\documentclass[prb,twocolumn,amsmath,amssymb,floatfix,footinbib,bibnotes,longbibliography]{revtex4-1}
\pdfoutput=1

\usepackage[colorlinks=true,citecolor=blue,linkcolor=magenta]{hyperref}
\usepackage{amsmath}
\usepackage{graphicx}
\usepackage{soul}
\usepackage{changepage}
\usepackage{fancyhdr}
\usepackage{amsthm,amssymb}
\usepackage{floatflt}
\usepackage{float}
\usepackage{array}
\usepackage[usenames,dvipsnames]{color}

\newcommand{\be}{\begin{align}}
\newcommand{\ee}{\end{align}}

\def \Tr{\mathrm{Tr}}

\def \be{\begin{equation}}
\def \ee{\end{equation}}
\def \ba{\begin{array}}
\def \ea{\end{array}}
\def \bea{\begin{eqnarray}}
\def \eea{\end{eqnarray}}

\def \ket#1{{|#1\rangle}}
\def \bra#1{{\langle#1|}}

\def \ba{\begin{align*}}
\def \ea{\end{align*}}

\def \mr{\mathrm}
\def \mb{\mathbf}

\def \mc{\mathcal}

\newcommand {\apgt} {\ {\raise-.5ex\hbox{$\buildrel>\over\sim$}}\ }
\newcommand {\aplt} {\ {\raise-.5ex\hbox{$\buildrel<\over\sim$}}\ }

\def \ba{\begin{align}}
\def \ea{\end{align}}

\def \mr{\mathrm}

\def \mc{\mathcal}

\begin{document}

\title{Exploration of the stability of many-body localization in $d>1$}
\author{Ionut-Dragos Potirniche,$^{1}$ Sumilan Banerjee,$^{2,3}$ and Ehud Altman$^{1,3}$ \\
\small \em 
$^{1}$ Department of Physics, University of California, Berkeley, CA 94720, USA\\
$^2$Department of Physics, Indian Institute of Science, Bangalore 560012, India\\
$^3$Department of Condensed Matter Physics, Weizmann Institute of Science, Rehovot 76100, Israel
}

\date\today

\begin{abstract}

Recent work by De Roeck \emph{et al.}~[Phys. Rev. B 95, 155129 (2017)] has argued that many-body localization (MBL) is unstable in two and higher dimensions due to a thermalization avalanche triggered by rare regions of weak disorder. To examine these arguments, we construct several models of a finite ergodic bubble coupled to an Anderson insulator of non-interacting fermions. We first describe the ergodic region using a GOE random matrix and perform an exact diagonalization study of small systems. The results are in excellent agreement with a refined theory of the thermalization avalanche that includes transient finite-size effects, lending strong support to the avalanche scenario. We then explore the limit of large system sizes by modeling the ergodic region via a Hubbard model with all-to-all random hopping: the combined system, consisting of the bubble and the insulator, can be reduced to an effective Anderson impurity problem. We find that the spectral function of a local operator in the ergodic region changes dramatically when coupling to a large number of localized fermionic states---this occurs even when the localized sites are weakly coupled to the bubble. In principle, for a given size of the ergodic region, this may arrest the avalanche. However, this back-action effect is suppressed and the avalanche can be recovered if the ergodic bubble is large enough. Thus, the main effect of the back-action is to renormalize the critical bubble size.
\end{abstract}

\maketitle

\section{Introduction}\label{sec:Introduction}

In a seminal paper~\cite{Basko2006}, Basko \emph{et al.} have argued that quantum systems evolving under their intrinsic dynamics can be many-body localized (MBL) and fail to thermalize in the presence of disorder and interactions (see also Ref.~\onlinecite{Gornyi2005}). This result has motivated a flurry of theoretical work exploring the nature of the many-body localized state and even a few recent experimental investigations~\cite{Schreiber2015,Smith2015}. The interest in this topic stems predominantly from the fact that MBL can protect quantum correlations from decoherence even at high energy densities and for arbitrarily long times (see Refs.~\onlinecite{Nandkishore2015review,Altman2015} and the references therein).

\begin{center}
\begin{figure}
\includegraphics[width=0.35\textwidth]{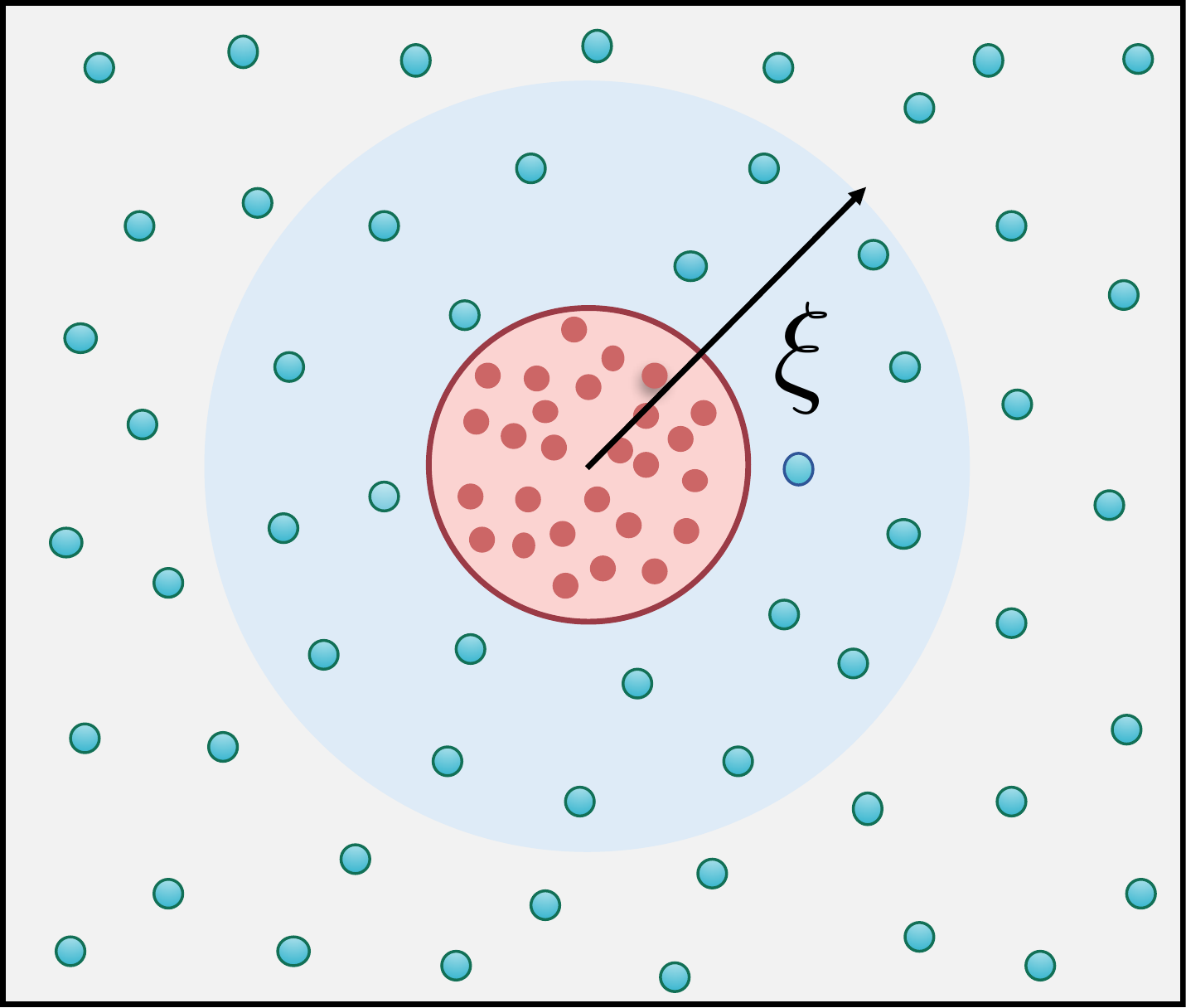}
\caption{Schematic illustration of an ergodic bubble inside an Anderson insulator. The red dots represent sites in the bubble, while the blue dots are the positions of the localized states of the insulator. The blue shaded region within a distance of a localization length $\xi$ from the bubble is strongly coupled to it.}
\label{fig.Bubble}
\end{figure}
\end{center}

At the same time, recent works have re-examined the case for the existence and stability of the MBL state~\cite{Imbrie2016,DeRoeck_mobility,DeRoeck2016}, going beyond the perturbative arguments of Basko \emph{et al}~\cite{Basko2006}. For instance, Imbrie has given a mathematical proof for the existence of MBL in spin chains with short-range interactions~\cite{Imbrie2016}. On the other hand, other analyses have pointed to non-perturbative effects that can \emph{destabilize} MBL under certain conditions~\cite{DeRoeck_mobility,DeRoeck2016}. In particular, De Roeck and Huveneers~\cite{DeRoeck2016} have argued that the MBL state is unstable in two or higher dimensions or in systems with interactions that decay sub-exponentially with distance. Their instability mechanism hinges upon finite ergodic ``bubbles'' of weak disorder that occur naturally inside an insulator (see Fig.~\ref{fig.Bubble}). According to this narrative, such rare regions may trigger a ``thermalization avalanche"---the avalanche commences by thermalizing the immediate surroundings of the bubble, thus creating a larger and more potent bubble which reinforces the process. Because this argument has far-reaching implications such as the absence of MBL in two-dimensional systems, it is important to test the crucial assumptions underlying this conclusion. 

In this paper, we scrutinize these assumptions using both an exact diagonalization study (ED) of small systems and a tractable toy model.

In the first part of the paper we describe the thermal bubble using a Gaussian orthogonal ensemble (GOE) random matrix in the Hilbert space of eight spins. To this core we gradually couple up to six more spins that, in the absence of coupling to the ergodic region, correspond to the local integrals of motion (LIOMs) of an Anderson insulator. The combined system, consisting of up to 14 spins, is diagonalized exactly.

\begin{center}
\begin{figure}
\includegraphics[width=\columnwidth]{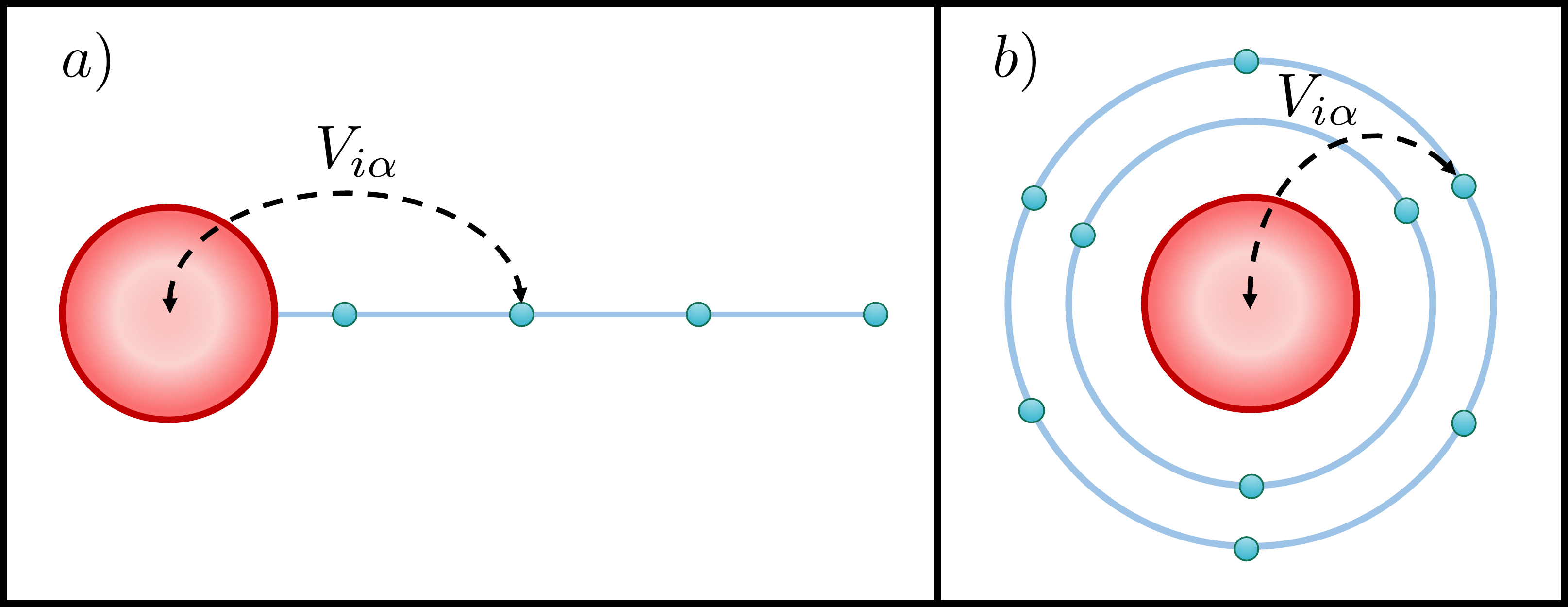}
\caption{The two geometries under consideration. In both cases, the coupling between the first LIOM and the bubble is set to be $V_1=1.0$, independent of the localization length $\xi$.\\
(a) One dimensional geometry: the other LIOMs, indexed by $\alpha \geq 2$, are located at distances $r_{\alpha} = (\alpha-1)a$ in units of $a=1$ from the ergodic bubble. The coupling strengths are $V_{\alpha} = V_1 e^{-r_{\alpha}/\xi}$ and $V_{\alpha} = V_{1}e^{-\sqrt{r_\alpha/\xi}}$ for the 1D insulator and 1D insulator with sub-exponentially decaying wave functions, respectively. (b) Two dimensional geometry: LIOMs are arranged in concentric layers around the ergodic bubble such that the $n^{\mathrm{th}}$ layer (for $n\geq 2$) is at a distance $r_n = (n-1)a$ from the bubble; this layer contains $n$ LIOMs indexed by $\alpha = \frac{n(n-1)}{2}+1,\dots,\frac{n(n+1)}{2}$ and their coupling strength is taken to be $V_{\alpha} = V_1 e^{-r_{n}/\xi}$.}
\label{fig:two_geometries}
\end{figure}
\end{center}

We study models with three different spatial structures of the coupling to represent (Fig.~\ref{fig:two_geometries}): (i) a one-dimensional Anderson insulator with an exponential decay of the localized single-particle wave functions; (ii) a one-dimensional insulator with a stretched exponential decay of the localized wave functions~\footnote{Even though this toy model is unphysical, we study it as an additional proxy for higher dimensional insulators since it allows us to analyze LIOMs farther away from the ergodic bubble than would be numerically possible in the case of a 2D insulator. We note, in passing, that it can also serve as a crude toy model for the eigenstate transition that separates the Ising-ordered and paramagnetic-MBL phases. Since the critical point is an infinite randomness point, the typical (though not average) correlations decay as a stretched exponential at the transition.}; (iii) a two-dimensional Anderson insulator.

In these systems we compute the eigenstate spectral function of a local operator acting on the farthest LIOM coupled to the thermal core which allows us to directly check if the LIOM is hybridized with the ergodic region~\footnote{LIOMs successfully hybridize with bath spins if and only if the Fermi Golden Rule decay rate of a local operator acting on the LIOM is non-zero.}. We then use direct probes of localization to test whether the fully coupled system is thermalizing and if the avalanche persists. For all three models, we find a quantitative agreement between the ED results and a refined theory of the thermalization avalanche~\cite{DeRoeck2016} that includes transient finite-size effects. Thus, the exact diagonalization study provides further support to the avalanche scenario. We note that these results go beyond previous numerical analyses in two main ways. First, we additionally investigate models in higher dimensions and in one dimension with longer-range coupling---models (ii) and (iii) above---in order to test for the existence of an inherent instability at any disorder strength. Second, we compare the numerical results with the predictions of a refined theory of the avalanche, which includes finite-size effects, to explore the physics of critical bubble sizes.

Although the numerical results for small systems are consistent with the avalanche scenario, we cannot rule out a failure of the instability in a much larger system. In the second part of the paper we discuss such a possible mode of failure caused by the collective back-action of the insulator onto the bubble. We describe the ergodic region using a Hubbard model of $N$ fermion sites with random, all-to-all hopping and on-site interactions. As before, to this core we couple non-interacting fermions that would otherwise realize a two-dimensional Anderson insulator (germane to the third model above). For large $N$, after averaging over the disorder in the bubble and in the bubble-insulator coupling, we show that the problem can be mapped to an Anderson impurity problem akin to the dynamical mean field theory (DMFT) approximation~\cite{Georges1996}. We compute the  spectral function of a local operator acting on the thermal core using an approximate impurity solver and track the evolution of the spectral function upon adding a growing number of Anderson sites localized at an increasing distance from the bubble (Fig.~\ref{fig.Bubble}). This allows us to assess the back-action of the localized region on the bath spectral function and, thus, test a key assumption of Ref.~\onlinecite{DeRoeck2016}. While the instability arguments hinge on having a weak back-action, we find a dramatic back-action effect even for reasonably weak bubble-insulator coupling. This cumulative---emergent---effect is caused by the many-body interactions between a very large number of weakly-coupled LIOMs and the ergodic bubble.

While the analyses of these solvable models point to a possible failure of the instability, there are two important caveats in this approach. First, the strong back-action is the result of quantum fluctuations induced by virtual hops of fermions from the Anderson insulator onto the interacting bubble. Thus, the effect is greatly suppressed when the surrounding insulator is strongly localized. Naively, this implies that the \emph{instability} of the insulator is more pronounced if the insulator is strongly, rather than weakly, localized. One possible resolution to this apparent paradox is that, although the solvable models allow us to compute the thermal spectral function, the quantity suitable for tracking and sustaining the instability is the spectral function in a \emph{typical} eigenstate~\cite{Serbyn2016}. Second, we emphasize that the thermal spectral function of an interacting system at non-zero temperatures cannot detect localization~\cite{Nandkishore2014a}. Thus, an open and interesting direction for future work would be the development of a different approach for assessing the avalanche effect in the case of a strong insulator.

The remainder of the paper is organized as follows. In Sec.~\ref{sec:Review_of_argument}, we briefly review the avalanche arguments of Ref.~\onlinecite{DeRoeck2016}. In Sec.~\ref{sec:ED}, we present the exact diagonalization study of a generic model for a random matrix bubble coupled to the Anderson insulator and we discuss various thermalization diagnostics of the instability argument. In Sec.~\ref{sec:HubbardAnderson}, we analyze a tractable model wherein we describe the ergodic region using a Hubbard model; this allows us to calculate thermal spectral functions for large system sizes that are inaccessible in ED studies. Finally, we summarize the results and outline possible extensions in Sec.~\ref{sec:Conclusions}.

\section{Review of the rare region instability argument}\label{sec:Review_of_argument}

Before describing our numerical results and toy models, we briefly summarize the main steps of the argument for the instability of MBL in the presence of a thermal bubble. The full system is described by the following Hamiltonian:
\begin{equation}\label{eq:RMTmodel}
\mc{H}=\mc{H}_b+\mc{H}_l+\mc{H}_{bl},
\end{equation}
where $\mc{H}_b$, describing the bubble, is a $2^N\times 2^N$ GOE random Hermitian matrix. For instance, $\mc{H}_b$ could correspond to a system of spinless fermions $(c_i^\dagger)$ on $i=1,\dots,N$ sites. The bubble has a typical many-body level spacing $\delta_b\sim \mc{W}/2^N$, where $\mc{W}$ is the many-body bandwidth, taken to scale linearly with the system size $N$~\footnote{In our numerics we take $\mc{W} = wN$, where $w \approx 2.6$.}.
According to the Eigenstate Thermalization Hypothesis (ETH)~\cite{Srednicki1994,Srednicki1999}, the thermal region is characterized by a smooth function $\rho(\omega)$ of the energy difference $\omega = E_n - E_m$ between energy eigenstates, $\ket{\Psi_n}$, of $\mc{H}_b$ such that the matrix element is given by $\langle \Psi_n|c_i^\dagger|\Psi_m\rangle\simeq \sqrt{\delta_b\rho(\omega)}\eta_{nm}$. Here, $\mc{H}_b|\Psi_n\rangle=E_n|\Psi_n\rangle$ and $\eta_{nm}$ is a random number with zero mean and unit variance. The function $\rho(\omega)$ is called the bubble spectral function (see Section~\ref{subsec:Thouless} for more details).

The insulating region is described in terms of a set of local integrals of motion (LIOM) $\{n_\alpha\}$, $\alpha=1,\dots,M$, via the Hamiltonian
\begin{align}
\mc{H}_l=\sum_\alpha \epsilon_\alpha n_\alpha, \label{eq.Hl}
\end{align}
 corresponding to an Anderson insulator of non-interacting fermions. We define $n_\alpha=\psi_\alpha^\dagger \psi_\alpha$, where $\psi_\alpha^\dagger$ is the operator creating a fermion in the localized eigenfunction $\phi_\alpha(\mb{r})\sim e^{-|\mb{r}-\mb{r}_\alpha|/2\xi}/\xi^{d/2}$ centered at $\mb{r}_\alpha$ in $d$ spatial dimensions. The LIOM energies obey $|\epsilon_\alpha|\leq W$ and $W$ is the single-particle bandwidth of the insulator. Envisaging the bubble as a ``quantum dot'' located at the origin, the bubble-insulator coupling is given by
\begin{align}\label{Eq:Hbl}
\mc{H}_{bl}=\sum_{i\alpha}(V_{i\alpha}c_i^\dagger \psi_\alpha+ \mr{h.c.}),
\end{align}
with $V_{i\alpha}\sim Ve^{-r_\alpha/\xi}$ decaying exponentially with the distance from the bubble. While a true MBL system contains interactions of the form $n_{\alpha}n_{\beta}$ between LIOMs, following Ref.~\onlinecite{DeRoeck2016}, we neglect them for the following two reasons. First, since the avalanche arguments focus on a potential instability of MBL, one normally considers ``the best-case scenario'' for localization, namely a single finite ergodic region in an otherwise non-interacting Anderson insulator. This follows from the expectation that interactions among LIOMs can only enhance the instability. Second, since we couple each LIOM to the random matrix bubble which renders the full system interacting, the sole effect of LIOM-LIOM coupling would be a higher-order, multi-spin, coupling between the LIOMs and the bubble.

The first step in the instability argument~\cite{DeRoeck2016} is to consider whether the LIOM closest to the quantum dot, namely the localized site with the strongest coupling $V_{i\alpha}$, gets hybridized with the bubble. The criterion for this is given by the following condition for the matrix element $\mc{T} \propto V|\langle \Psi_n|c_i^\dagger|\Psi_m\rangle|\approx V\sqrt{\delta_b\rho(W)}\gg \delta_b$, where $|E_n-E_m|\approx W$. Crucially, this entails a non-zero Fermi Golden Rule (FGR) decay rate $\sim \mc{T}^2/\delta_b$ for the LIOM. If this is the case, then the site can be considered as part of the bubble and, once again, we assume the ETH ansatz for the matrix elements. Since the LIOM is absorbed into the ergodic region, then the Hilbert space dimension of the combined system increases by a factor of 2, i.e. the level spacing gets reduced to $\delta_E\simeq \delta_b/2$. However, in the process, the spectral function of the ergodic grain also gets modified from $\rho(\omega)$ to $\tilde{\rho}(\omega)$. Based on certain assumptions about the eigenfunctions $|\widetilde{\Psi}_n\rangle$ of the combined system as a linear superposition of product states $\{\ket{\Psi_n}\otimes\ket{n_\alpha}\}$ of the initial bubble and the LIOM, Ref.~\onlinecite{DeRoeck2016} argues that $\tilde{\rho}(\omega)\simeq \rho(\omega)$, i.e. that the spectral function remains essentially the same after the LIOM becomes a \emph{bona fide} member of the bubble. 

If we iterate this argument multiple times such that the bubble grows to a radius $R$ (see Fig.\ref{fig.Bubble}), then the level spacing becomes $\delta_E(R)\sim \delta_b e^{-A_dR^d}$, where $A_d\sim \mc{O}(1)$. The matrix element to absorb an additional LIOM at distance $R$ is $\mc{T}_R\sim V\sqrt{\delta_E(R)\rho(W)}e^{-R/\xi}$, leading to the following condition for the hybridization of the site into the bubble:
\begin{equation}\label{eq:hybridization_condition}
\frac{\mc{T}_R}{\delta_E(R)}\sim V \sqrt{\frac{\rho(W)}{\delta_b}} e^{A_dR^d/2-R/\xi}\gg 1.
\end{equation}
Evidently, the above condition can always be satisfied for $R\to \infty$ in $d>1$, rendering the localized state unstable in any dimension higher than one. However, note that the exponential has a minimum at $R^{*} = \left[2/(dA_d\xi\right)]^{1/(d-1)}$ when $d>1$. For the avalanche to continue indefinitely, it must survive around $R^{*}$. Thus, a necessary and sufficient condition for the LIOM at an arbitrary distance $R>R^{*}$ to be hybridized with the ergodic region is that $\mc{T}_{R^{*}}/\delta_E(R^{*}) \gg 1$. This immediately holds if the initial bubble size $N> N^*\sim \left(1/\xi\right)^{d/(d-1)}$. For $d=1$, the function in Eq.~\ref{eq:hybridization_condition} is monotonically increasing or decreasing if $\xi> \xi_c$ or $\xi< \xi_c$, respectively, where $\xi_c = \frac{2}{A_d}$.

Lastly, we want to emphasize the two key assumptions embedded in the instability argument. First, the thermalization avalanche continues indefinitely if and only if the FGR decay rate for all LIOMs is non-zero. Second, whenever a LIOM is hybridized with the bubble, the ETH holds and the spectral function of a local operator acting on the ergodic region stays qualitatively the same.

\section{Exact diagonalization of a generic model}\label{sec:ED}

We now test these two assumptions in an exact diagonalization study of a model akin to the one considered in Ref.~\onlinecite{DeRoeck2016}  and defined in Eq.~\ref{eq:RMTmodel}. For computational simplicity, we take $\mathcal{H}_{b}$, describing the ergodic bubble centered at the origin, to be a $2^N \times 2^N$ GOE random matrix on spin degrees of freedom $\sigma_{i}^{\{x,y,z\}}$ (where $i =1,\dots,N$) such that the many-body bandwidth scales linearly with $N$. We then fix the size of the ergodic region to $N=8$ spins. We characterize the LIOMs also using Pauli spin variables $\tau^{\{x,y,z\}}_{\alpha}$, where $\alpha=1,\dots,M$, such that $\mathcal{H}_l = \sum_{\alpha} \epsilon_\alpha \tau_{\alpha}^z$. The random fields $\epsilon_{\alpha}$ are sampled from the uniform distribution on $[0.5,1.5]$. 

The geometries we consider are shown in Fig.~\ref{fig:two_geometries}. The bubble-insulator coupling is taken to be $\mathcal{H}_{bl} = \sum_{\alpha} V_{\alpha} \sigma_{1}^{x}\tau_{\alpha}^{x}$, where $V_{\alpha}$ depends on the geometry under consideration (see Fig.~\ref{fig:two_geometries}) and we set the coupling of the first LIOM to be $V_1 = 1.0$, independent of $\xi$. Since the bubble is described by a random matrix without any notion of locality, coupling all LIOMs to the same bubble operator $\sigma_1^x$ does not affect the subsequent discussion.

\begin{center}
\begin{figure}
\includegraphics[width=\columnwidth]{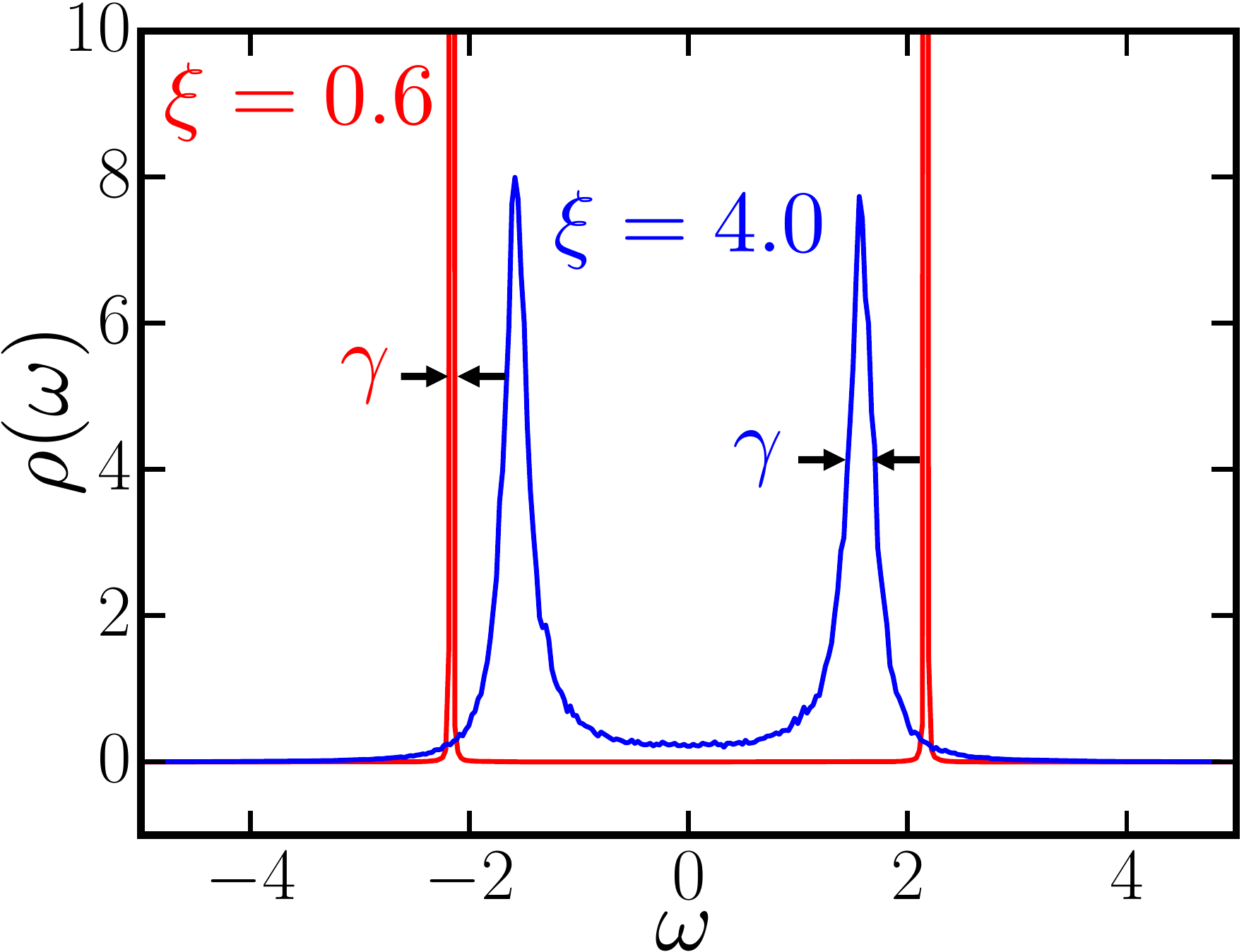}
\caption{The spectral function $\rho(\omega)$ of a local operator acting on the $M^{\mathrm{th}}$ LIOM in a fixed disorder realization $\{\epsilon_{\alpha}\}$ and averaged over $N_A$ eigenstates in the middle of the band.
$\rho(\omega)$ is plotted as histogram with a bin width equal to twice the many-body level spacing, $2\delta_E$. The spectral function exhibits two peaks located at $\pm 2\epsilon_{M}$. For a large localization length $\xi$, the width $\gamma$ of these peaks is much larger than the level spacing, indicating that the LIOM is hybridized with the ergodic bubble. For a small localization length, the LIOM is not hybridized and the width of the peaks is limited by the level spacing.}
\label{fig:Spectral_two_xis}
\end{figure}
\end{center}

\begin{figure*}[ht]
\centering
\includegraphics[width=\textwidth]{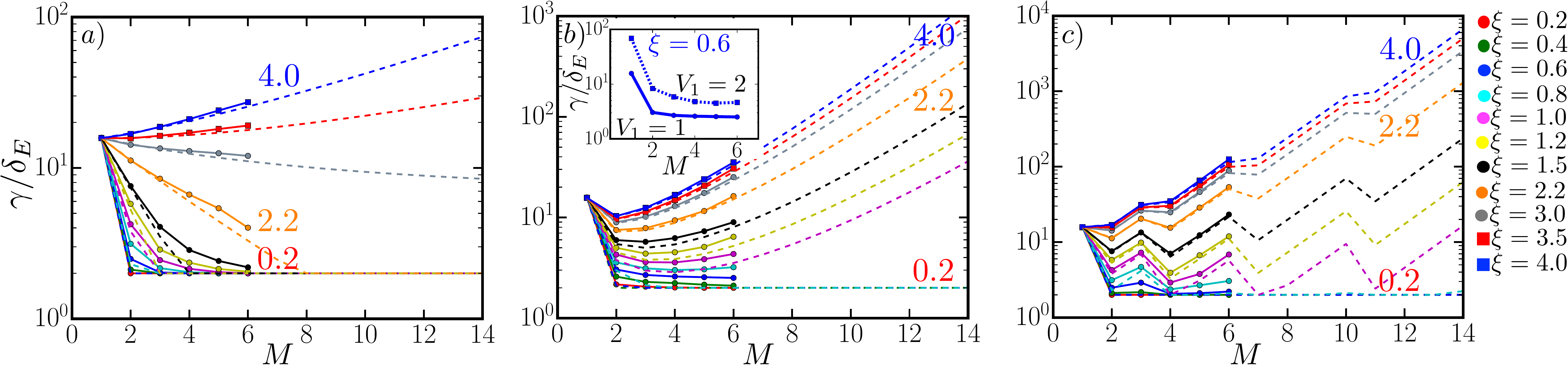}
\caption{Exact diagonalization (ED) results for the spin model defined in Section~\ref{sec:ED} using an ergodic bubble of $N=8$ spins coupled to $M=1,\dots,6$ LIOMs. We plot the ratio $\gamma/\delta_E$ between the width $\gamma$ of the spectral function peaks of a local operator acting on the $M^{\mathrm{th}}$ LIOM (see Fig.~\ref{fig:Spectral_two_xis}) and the many-body level spacing $\delta_E$ as a function of the number of LIOMs added. Different curves and colors (online) correspond to different localization lengths in the bulk: the solid curves represent numerical ED results, whereas the dashed lines represent the theoretical expectations from Eqs.~\ref{eq:ETH_width_expectation} and~\ref{eq:ETH_width_expectation_2} based on ETH assumptions. (a) One-dimensional geometry with exponentially decaying couplings between the bubble and LIOMs. (b) One-dimensional geometry with a stretched exponential decay of couplings. (c) Two-dimensional geometry with an exponential decay of couplings. In (b) and (c) the failure of the avalanche is due to the sub-critical bubble size. In these cases, the avalanche can be restored by increasing the bubble size $N$ or by increasing the bare bath-LIOM coupling $V_1$, as verified in the inset of panel (b) for the one-dimensional model with stretched exponentials.}
\label{fig:Spectral}
\end{figure*}

\subsection{LIOM spectral functions}\label{subsec:LIOM_spectral}

We obtain the full spectrum and many-body eigenstates $\mathcal{H}\ket{\Psi_n} = E_{n} \ket{\Psi_n}$ through the exact diagonalization (ED) of the spin Hamiltonian described in the preceding paragraphs. To check whether the LIOMs are successfully hybridized with the bubble spins, we define the spectral function of a local operator $\tau_{M}^{x}$ acting on the LIOM farthest from the ergodic region for a given eigenstate $|\Psi_n\rangle$ as
\begin{equation}\label{eq:single_spectral}
\rho_{n} (\omega) = 2\pi \sum_{m\neq n} \left|\bra{\Psi_n}\tau_{M}^{x}\ket{\Psi_{m}}\right|^2 \delta(\omega - \omega_{mn}),
\end{equation}
where $\omega_{mn} = E_{m}-E_{n}$. We can also think of $\rho_n(\omega)$ as the dynamical structure factor $\mathrm{Im} \int_{0}^{\infty} e^{i\tilde{\omega}t} \bra{\Psi_n}\tau_M^{x}(t)\tau_{M}^{x}(0)\ket{\Psi_{n}}dt$ where $\tilde{\omega} = \omega + i0^{+}$. Note that it also obeys a sum rule: $\int_{-\infty}^{+\infty} \rho_{n}(\omega)d\omega = 2\pi(1-\bra{\Psi_n}\tau_{M}^x\ket{\Psi_n}^2)\approx 2\pi$.

In a fixed disorder realization for the LIOM fields $\{\epsilon_{\alpha}\}$, the eigenstate spectral function $\rho_n(\omega)$ exhibits two peaks located at $\pm 2\epsilon_{M}$, where $1\leq 2|\epsilon_M|\leq 3$, as shown in Fig.~\ref{fig:Spectral_two_xis}. This is due to the fact that flipping the LIOM farthest from the bubble requires an energy of approximately $2\epsilon_{M}$. We average over $N_A = 2^{N+M}/(N+M)$ eigenstates in the middle of the band where the putative ergodicity is the most robust: $\rho(\omega) = N_A^{-1}\sum_{n}\rho_{n}(\omega)$. Because the LIOMs do not interact directly with each other, the position of the peak is not expected to depend on the eigenstate, but only on the value of the field $\epsilon_M$ on this LIOM.

We can diagnose the extent to which the farthest LIOM is hybridized with the bath by looking at the width $\gamma$ of the spectral peaks of $\rho(\omega)$ compared to the many-body level spacing---for each disorder realization we compute the ratio $\gamma/\delta_E$ between the spectral width $\gamma$ and the numerically computed many-body level spacing $\delta_E$. We then study the evolution of the disorder averaged $\overline{\gamma/\delta_E}$~\footnote{We take $N=8$ and average over 5000, 4000, 3000, 2000, 1000, 1000, and 200 disorder realizations for $M = 0,1,2,3,4,5,$ and $6$ LIOMs, respectively.} as a function of $M$, the number of LIOMs coupled to the ergodic bubble, as shown in Fig.~\ref{fig:Spectral} for the three geometries under consideration. 

First, we notice that for very small localization lengths $\xi$ the ratio $\overline{\gamma/\delta_E}\approx 2$ which means that $\gamma \sim e^{-\mathcal{O}(M)}$. In particular, this entails that the FGR decay rate for the $M^{\mathrm{th}}$ LIOM is approximately zero and the many-body eigenstates of the fully coupled system are product states of the form $\ket{\Psi_n}\otimes \ket{\uparrow_{M}}$ and $\ket{\Psi_n}\otimes \ket{\downarrow_{M}}$ with energies $E_n+ \epsilon_{M}$ and $E_n-\epsilon_{M}$, respectively. Thus, $\rho(\omega)$ consists of two delta functions located at $\pm 2\epsilon_{M}$. Since we compute spectral functions as energy histograms wherein the bin size is taken to be $2\delta_E$, this means that a ``delta function'' peak has a width $\gamma/\delta_{E} = 2$. We have confirmed that the farthest LIOM does not get hybridized with the ergodic bubble by checking that the entanglement entropy of this LIOM is zero (see the Supplementary Material~\cite{Supplementary} for details).

Second, for a large localization length $\xi$ we find that the peaks centered around $\pm 2\epsilon_{M}$ get broadened such that $\overline{\gamma/\delta_E}\gg 1$ (see Fig.~\ref{fig:Spectral}). This is due to the fact that now there are many accessible multi-spin processes such as flipping the LIOM by flipping bubble spins and absorbing energy from the ergodic bubble. In other words, the LIOMs are successfully hybridized with the bubble, the full many-body eigenstates are superpositions of many product states of the form $\ket{\Psi_n}\otimes \ket{\tau_{M}^{z}}$, and the FGR decay rate is non-zero.

We now compare these numerical results with the avalanche scenario~\cite{DeRoeck2016}. First, by the above arguments, we note that if the FGR is violated, then $\gamma/\delta_E = 2$. Second, if the FGR decay rate for the $M^{\mathrm{th}}$ LIOM is non-zero, then the width $\gamma =\gamma_M\sim \frac{V_{M}^2}{w}$, where $w = \mc{W}/(N+M-1)$ and $V_{M}$ is the coupling strength between the farthest LIOM and the bubble: (i) for the $d=2$ geometry $V_{M} = V_{1}e^{-r_M/\xi}$, where $r_M = (n-1)$ if the $M^{\mathrm{th}}$ LIOM is located on the $n^{\mathrm{th}}$ layer; (ii) for the $d=1$ geometry with exponentially decaying LIOM wave functions $V_{M} = V_{1}e^{-r_M/\xi}$, where $r_M = (M-1)$; (iii) for the $d=1$ geometry with sub-exponentially decaying wave functions $V_{M} = V_{1}e^{-\sqrt{r_{M}/\xi}}$, where $r_{M} = (M-1)$. Lastly, the many-body level spacing is $\delta_{E} = w(N+M-1)/2^{N+M-1}$.

With these expressions in hand, we expect that
\begin{align}\label{eq:ETH_width_expectation}
\nonumber \frac{\gamma}{\delta_{E}} =  \max \Bigg[2,  & \frac{\gamma_1}{\delta_b}\frac{N}{N+M-1} \times \\
 &\times \exp\left((M-1)\log 2 - 2\frac{r_M}{\xi}\right)\Bigg]
\end{align}
for the two models with exponentially decaying wave functions: see the dashed curves in Fig.~\ref{fig:Spectral}(a) and Fig.~\ref{fig:Spectral}(c) for $d=1$ and $d=2$, respectively. The maximum in Eq.~\ref{eq:ETH_width_expectation} ensures that the ratio does not drop below 2 in the case where the FGR is violated. Note that when the FGR holds, we have written the ratio $\gamma_M/\delta_E(M)$ as a function of $\gamma_{1}/\delta_b$ to ensure that both the numerical and theoretical curves have the same starting point, namely the same ratio for the first LIOM that is coupled. The first term in the exponential in Eq.~\ref{eq:ETH_width_expectation} comes from the many-body level spacing, whereas the second one comes from the exponential decay of the coupling strength $V_M$. The logarithmic correction (in $M$) comes from the linear scaling of the many-body bandwidth.

Note that for the 1D model with a stretched exponential decay of the localized wave functions [Fig.~\ref{fig:Spectral}(b)], the expression becomes
\begin{align}\label{eq:ETH_width_expectation_2}
\nonumber \frac{\gamma}{\delta_E} = \max \Bigg[2,  & \frac{\gamma_1}{\delta_b}\frac{N}{N+M-1}\times \\
&\times \exp\left((M-1)\log 2 - 2\sqrt{\frac{r_M}{\xi}}\right)\Bigg].
\end{align}

The comparison between the above expectations for $\gamma/\delta_E$ and the numerically obtained $\overline{\gamma/\delta_E}$ yields an excellent agreement for all three models, as shown in Fig.~\ref{fig:Spectral}. 
We emphasize the monotonic behavior of the curves in the case of the 1D model with an exponential decay of the LIOM wave functions: both the numerical results and the avalanche scenario predict that $\gamma/\delta_E$ is monotonically increasing when $\xi>\xi_c = \frac{2}{\log 2}$ and decreasing when $\xi<\xi_c$---this suggests that there is a localization-delocalization phase transition as a function of $\xi$ independent of the initial bubble size~\cite{DeRoeck2016}. In contrast to this behavior, the curves for the 2D model and the 1D model with stretched exponentials behave \emph{non-monotonically} with a minimum at $M^{*}$. This suggests that for any given $\xi$, a sufficiently large initial bubble will thermalize the whole system.

We now test this latter point, to wit, that the apparent ``localizing behavior'' at small $\xi$'s in Fig.~\ref{fig:Spectral}(b) and Fig.~\ref{fig:Spectral}(c) is due to an insufficiently potent initial thermal bubble. While it is numerically prohibitive to increase the bubble size $N$, we can increase $V_1$, the coupling strength to the bubble. We set $V_{1}=2 < w=2.6$ and keep $N=8$ fixed for the 1D model with a stretched exponential decay of the couplings. As shown in the inset of Fig.~\ref{fig:Spectral}(c), the results are still in excellent quantitative agreement with the avalanche scenario and the predictions from Eq.~\ref{eq:ETH_width_expectation_2}: for a fixed localization length $\xi = 0.6$, the ergodic region is more effective at thermalizing the LIOMs than in the case of $V_1=1$ since the $\gamma/\delta_E(M)$ curve is shifted upwards. 

\begin{figure*}[ht]
\centering
\includegraphics[width=\textwidth]{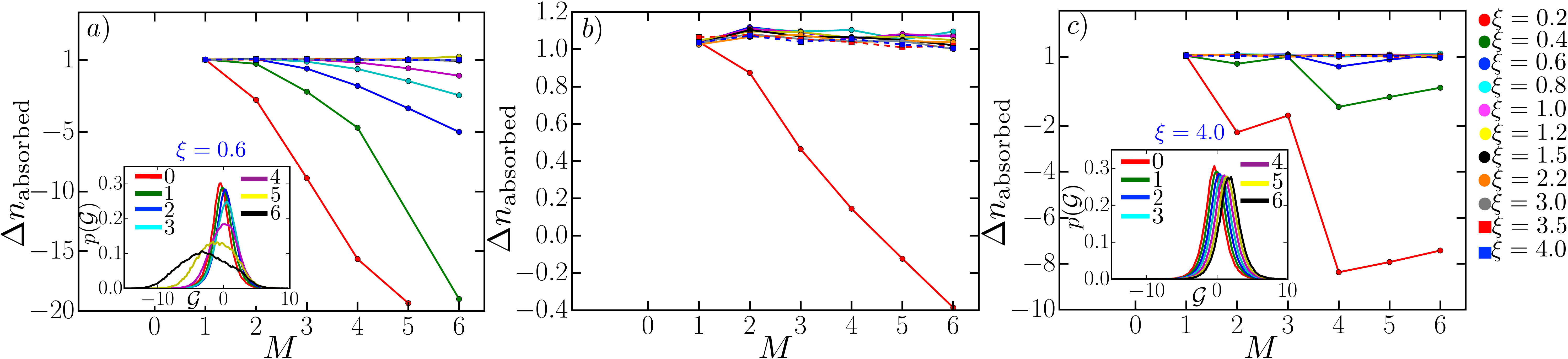}
\caption{Increase of the effective bath size due to adding the $M^{\text{th}}$ LIOM as measured by the change of the Thouless parameter (see Eq.~\ref{Eq:nabsorbed}).  (a) One-dimensional geometry with exponentially decaying couplings. The different curves and colors (online) correspond to different localization lengths: as described in the main text, a transition occurs at $\xi_c = 2/\log 2 \approx 2.9$. The inset shows how the distribution $p(\mathcal{G})$ for $\xi=0.6$ evolves with the number of added spins $M$. The distribution eventually collapses to the left and broadens, signaling localization.
(b) One-dimensional geometry with stretched exponential decay of interactions.
(c) Two-dimensional geometry with exponentially decaying couplings. Inset: evolution of the distribution $p(\mathcal{G})$ for $\xi=4.0$ in the one-dimensional geometry with exponentially decaying interactions. The distribution moves to the right at a constant rate, signaling thermalization.}
\label{fig:Thouless}
\end{figure*}

Thus, after accounting for transient finite-size effects and the bare coupling strengths, the avalanche scenario is found to be in remarkably good quantitative agreement with our numerics. As long as LIOMs are hybridized with the ergodic region, the fully coupled system is thermal and the avalanche persists indefinitely. It only stops when the hybridization fails, but a larger or more potent ergodic region can overcome this issue.

\subsection{Bath spectral functions}\label{subsec:Thouless}

Above, we have explored the conditions under which LIOMs added incrementally to a finite bath successfully hybridize with it in the usual sense of Fermi's Golden Rule: to wit, the local spectral function of a LIOM is broader than the many-body level spacing. An important assumption in the arguments of Ref.~\onlinecite{DeRoeck2016} is that once a LIOM is hybridized with the bath in this way, it is fully absorbed into it. In other words, the low-frequency characteristics of the bath after absorbing the LIOM should be the same as those of a random matrix with one additional degree of freedom. 

We test this assumption by analyzing the spectral function of an original bath spin. Specifically, we consider the many-body ``Thouless conductance'' defined in Ref.~\onlinecite{Serbyn2015} (it will also be defined below for completeness). The typical value of this parameter in a thermalizing system is expected to grow linearly with the size of the bath, with a slope equal to the entropy density.

The computational scheme is as follows. After obtaining the full spectrum and eigenstates, we apply a local perturbation $\mathcal{O} = \sigma_{N}^{z}$ on a bubble spin different than the one to which we have coupled the LIOMs (i.e. $\sigma_{1}$) to mitigate the severity of finite-size effects~\footnote{We have checked that our results are consistent for a different local perturbation $\mathcal{O}_2 = \sigma_{N}^{z}\tau_{1}^{z}$.}. We rearrange the eigenstates based on the perturbed energies $E_{n}' = E_{n} + \bra{\Psi_n}\mathcal{O}\ket{\Psi_{n}}$ and compute the matrix element between nearby eigenstates $\mathcal{T}_{n,n+1} = \bra{\Psi_{n}}\mathcal{O}\ket{\Psi_{n+1}}$. Then we define the Thouless parameter as in Ref.~\onlinecite{Serbyn2015}:
\begin{equation}
\mathcal{G} = \log \frac{|\mathcal{T}_{n,n+1}|}{E_{n+1}'-E_{n}'}.
\end{equation}
We compute this parameter for $N_A=2^{N+M}/(N+M)$ eigenstates in the middle of the band and for many disorder realizations~\footnote{We average over 5000, 4000, 3000, 2000, 1000, 1000, and 200 disorder realizations for $M = 0,1,2,3,4,5,$ and $6$ LIOMs, respectively.} to produce a distribution $p(\mathcal{G})$. 

The results for different values of the localization length $\xi$ are shown in Fig.~\ref{fig:Thouless}. For a thermalizing quantum system whose eigenstates obey the ETH, we expect $\mathcal{G} \sim (N+M)$ since the matrix element $\mathcal{T}_{n,n+1}$ is essentially an overlap between two random states in a $2^{N+M}$-dimensional Hilbert space and the level spacing is $\delta_E \sim 2^{-(N+M)}$. Conversely, if the system is localized then two nearby eigenstates can be connected only via extensively many rearrangements of $(N+M)$ local integrals of motion and $\mathcal{G} \sim -(N+M)$, as shown in Ref.~\onlinecite{Serbyn2015}. 

\begin{figure*}[ht]
\centering
\includegraphics[width=0.9\textwidth]{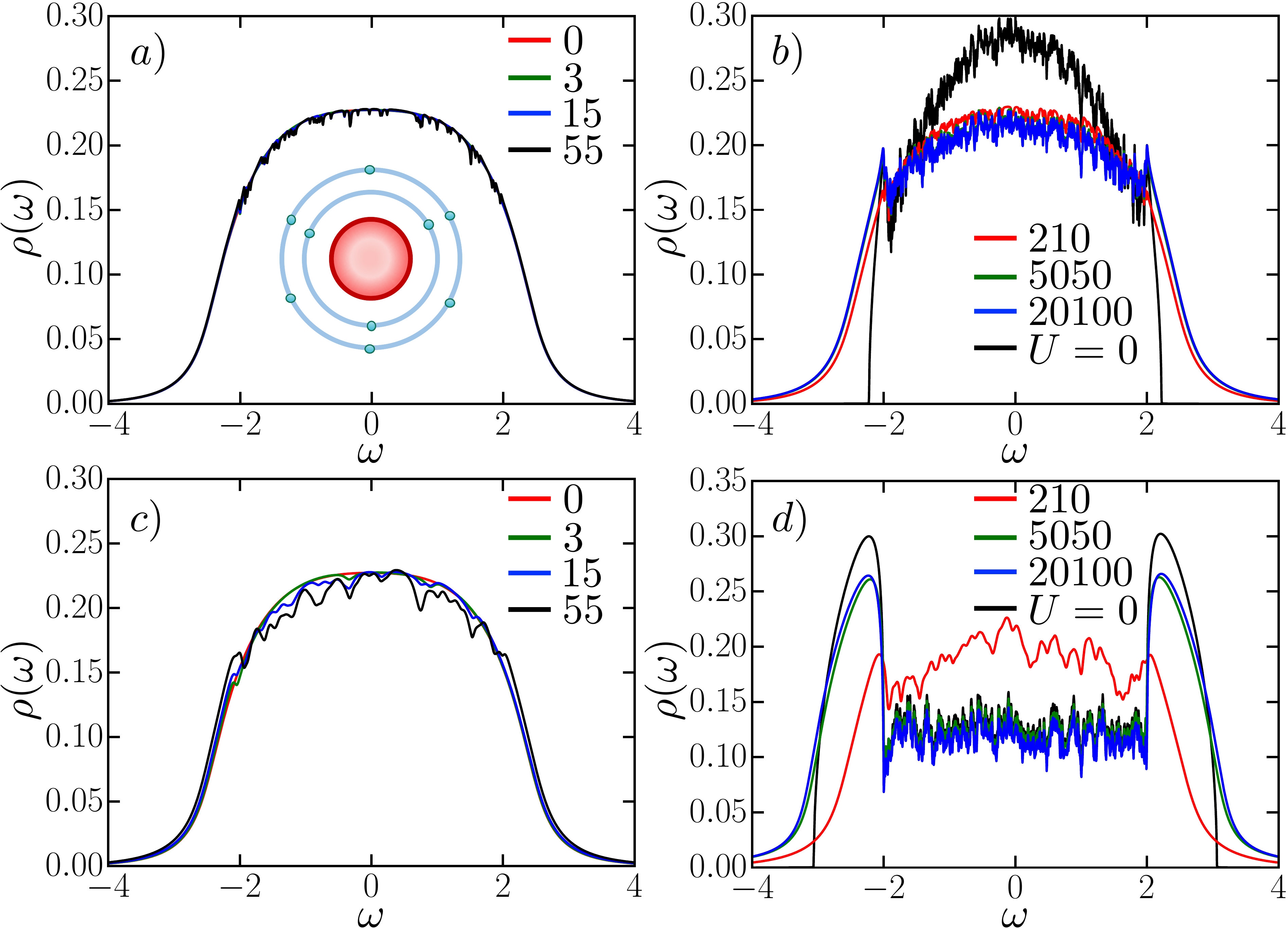}
\caption{The evolution of the bubble spectral function $\rho(\omega)$ for (a)--(b) $V=0.5$ and (c)--(d) $V=1.5$ as a function of $M$, the number of LIOMs coupled to a Hubbard bubble of $N=30$ sites. Each curve (color) corresponds to a different $M$. The inset in (a) shows the 2D geometry in which concentric layers of LIOMs are coupled to the bubble. Here, $t=1$, $U=2$, $W=2$, $\xi=10$, $n_L=1$, and $T=2$. In (b), for the weak coupling $V=0.5$, $\rho(\omega)$ does not change much even at large $M$. For comparison, we also show $\rho(\omega)$ for the non-interacting case ($U=0$). For the stronger coupling $V=1.5$, $\rho(\omega)$ changes drastically in (c) and (d) when large numbers of LIOMs are coupled to the bubble. In (d), $\rho(\omega)$ for the interacting case becomes identical to the non-interacting one over an interval $-W<\omega<W$ for large $M$.}
\label{fig.HubbardAnderson}
\end{figure*}

Our numerical results support the assumption that when a LIOM is hybridized according to the local golden rule criterion, then that LIOM is truly an extra bath degree of freedom. To check this, we numerically compute 
\begin{equation}\label{Eq:nabsorbed}
\Delta n_{\mathrm{absorbed}} = \frac{2}{\log 2}\left[\langle\mathcal{G}(M+1)\rangle-\langle\mathcal{G}(M)\rangle\right],
\end{equation}
where $\langle\mathcal{G}(M)\rangle$ is the average value of the Thouless parameter when we have coupled $M$ LIOMs. If $\Delta n_{\mathrm{absorbed}} = 1$, then the Thouless parameter increases as it would in the case of adding exactly one more strongly coupled degree of freedom to the bath. This is illustrated in Fig.~\ref{fig:Thouless}. 

On the other hand, in situations where ETH arguments predict the failure of the avalanche (either due to a sub-critical bubble size or a sub-critical localization length), adding LIOMs leads to a sublinear increase or even decrease of $\mathcal{G}$. In this case, each additionally coupled LIOM contributes as less than a full degree of freedom to the bath. In this situation, the distribution $p(\mathcal{G})$ of the Thouless parameter also becomes broader with each additional LIOM (inset of Fig.~\ref{fig:Thouless}).

\begin{figure*}[ht]
\centering
\includegraphics[width=0.9\textwidth]{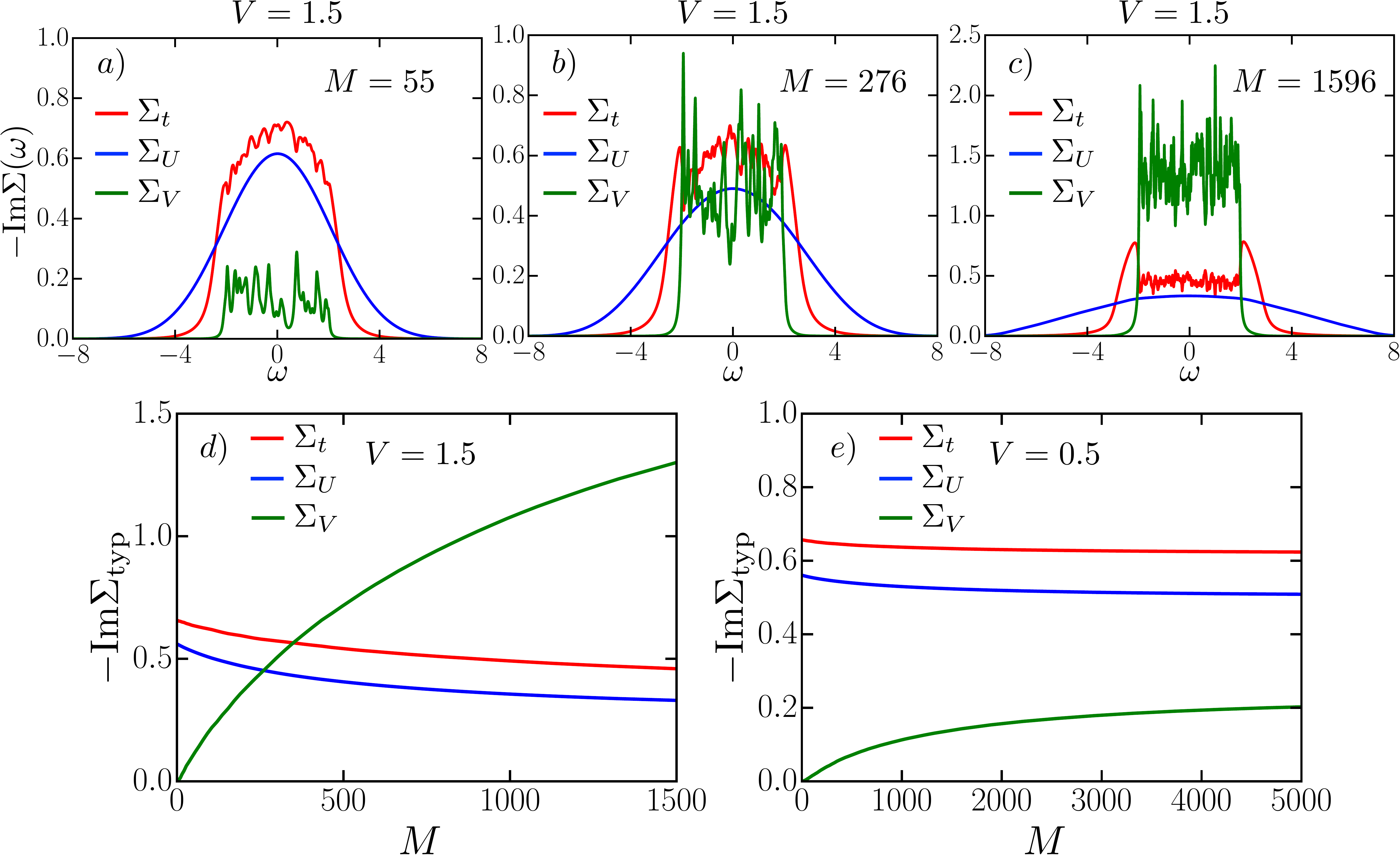}
\caption{(a)--(c) The imaginary part of the three different contributions, $\Sigma_t$ (due to the hopping), $\Sigma_U$ (due to the interactions in the bubble), and $\Sigma_V$ (due to the bubble-LIOM coupling), in the self-energy of the bubble for $V=1.5$ at three different values of $M$. We find that $\Sigma_V$ dominates the other two self-energies at large $M\apgt 300$. (d) and (e) compare the typical values (see the main text for the definition) of the three contributions for $V=1.5$ and $V=0.5$, respectively. For $V=1.5$ in (d), $\Sigma_V$ crosses $\Sigma_t$ and $\Sigma_U$ as a function of $M$, signaling the onset of the strong back-action in the spectral function, as shown in Fig.~\ref{fig.HubbardAnderson}(d).}
\label{fig.SelfEnergies}
\end{figure*}

We reiterate that even this eventual termination of the avalanche is predicted by the avalanche scenario---it is due to the violation of the FGR and the failure of the added LIOMs to hybridize with the bath. In the 1D model with exponentially localized LIOM wave functions, this signals a true localization transition~\cite{DeRoeck2016} tuned by the localization length $\xi$. However, in the 2D model and the 1D model with stretched exponential interactions the termination of the avalanche is \emph{not} a phase transition and it can be avoided by starting with a larger or more potent initial thermal bubble. Finally, as discussed in the Supplementary Material~\cite{Supplementary}, the non-monotonic behavior of $\langle\mathcal{G}\rangle$ at small localization lengths stems from a non-trivial variation of the bath spectral function as $M$ increases due to the breakdown of typicality.

\section{A model for back action on the bubble}\label{sec:HubbardAnderson}

The exact diagonalization study described above provides evidence in support of the avalanche scenario. But there is a key assumption in Ref.~\onlinecite{DeRoeck2016} that we have not yet explicitly checked, namely that the spectral function $\rho(\omega)$ of a local operator acting on the ergodic bubble does not change qualitatively as an ever-increasing number of LIOMs are hybridized with the bubble degrees of freedom. This can be particularly problematic if the number of added LIOMs is very large. In particular, we ask if there is a collective effect due to the coupling of many LIOMs which we cannot capture by adding a few LIOMs one by one. Such an effect is inaccessible in the exact diagonalization approach taken in the previous sections.

To explore this regime, we introduce a model of an ergodic bubble coupled to an Anderson insulator which admits a controlled treatment of the interaction effects in the bubble. Once again, we consider a Hamiltonian that can be written as a sum of three parts: $\mathcal{H}_b$, $\mathcal{H}_l$, and $\mathcal{H}_{bl}$ as in Eq.~\ref{eq:RMTmodel}. The ergodic bubble is described by a Hubbard model Hamiltonian of interacting fermions hopping on $N$ sites
\begin{align}
\mathcal{H}_b= & \frac{1}{\sqrt{N}}\sum_{ij\sigma}t_{ij,\sigma}c_{i\sigma}^{\dagger}c_{j\sigma}+U\sum_{i}n_{i\uparrow}n_{i\downarrow}, 
\label{eq.Hubbard}
\end{align}
where the single-particle hopping matrix elements $t_{ij,\sigma}=t_{ji,\sigma}^*$ are sampled from a GOE random matrix with $\overline{|t_{ij,\sigma}|^2}=t^2$. The normalization $1/\sqrt{N}$ ensures the proper thermodynamic limit at large-$N$. 

The LIOMs sitting outside the bubble are described by Eq.~\ref{eq.Hl} and their energies $\epsilon_\alpha\in [-W,W]$ are sampled from a uniform random distribution. We consider a 2D geometry of concentric circles, as shown in the inset in Fig.~\ref{fig.HubbardAnderson}(a), such that the number of insulating sites residing on a layer of radius $r$ is proportional to $r$. This ensures that the total number of insulating sites $M$ grows as $r^2$. In particular, we take $r_\alpha=r_0 (n-1)$ for the LIOMs on the $n$-th circle and $r_0=1/(\sqrt{2\pi n_{L}})$ is a length scale related to the areal density $n_{L}$ of the LIOMs.

Finally, the coupling of the LIOMs to the bubble degrees of freedom is given by
\be
\mc{H}_{bl}=\sum_{i\alpha \sigma}(V_{i\alpha,\sigma}c_{i\sigma}^\dagger \psi_\alpha+\mathrm{h.c.}).
\label{eq:V}
\ee
Note that this is different from the coupling considered earlier in Eq.~\ref{Eq:Hbl} since every LIOM couples to every degree of freedom in the bubble, not just to a single bath degree of freedom. Also, unlike in Eq.~\ref{Eq:Hbl}, $V_{i\alpha}$ are random complex numbers chosen from a Gaussian distribution with zero mean and  $\overline{|V_{i\alpha}|^2}=V_{\alpha}^2/2\sqrt{N}$, where $V_\alpha^2=(V/\xi)^2e^{-r_\alpha/\xi}$. The factor of $1/\xi$ in $V_\alpha$ arises due to the scaling $\phi_\alpha(\mathbf{r}) \propto 1/\xi$ of the single-particle LIOM eigenfunctions in 2D. Secondly, since the underlying microscopic coupling is \emph{local}, the contribution of $\mc{H}_{bl}$ to the free energy must scale as the surface area of the bubble, namely as $N^{(d-1)/d}$. The particular scaling of $\overline{|V_{i\alpha}|^2}$ with $N$ ensures that the contribution of the coupling term~\eqref{eq:V} in the action (see the last term in Eq.~\ref{eq:Action} below) scales as $\sqrt{N}$ in $d=2$.

In what follows, we average over the random hopping $t_{ij}$ and the bubble-LIOM coupling $V_{i\alpha,\sigma}$ using replicas in a fixed realization of the LIOM on-site energies $\{\epsilon_\alpha\}$. This averaging procedure over the couplings associated with the bubble degrees of freedom assumes that the self-averaging property holds for an ergodic bubble with a large number of sites $N$. The replicated partition function after disorder averaging is $\overline{Z^n}=\int \mc{D}(\bar{c}_a,c_a,\bar{\psi}_a,\psi_a)e^{-S[\bar{c}_a,c_a,\bar{\psi}_a,\psi_a]}$, where the imaginary-time action is
\begin{align}
S&=\int_0^\beta d\tau \left[\sum_{ia}\left(\sum_\sigma\bar{c}_{i\sigma a}(\tau)\partial_\tau c_{i\sigma a}(\tau)-Un_{i\uparrow a}(\tau)n_{i\downarrow a}(\tau)\right) \right.\nonumber \\
&\left.+\sum_{\alpha a}\bar{\psi}_{\alpha a}(\tau)(\partial_\tau+\epsilon_\alpha)\psi_{\alpha a}(\tau) \right]+\frac{N}{2}\int d\tau d\tau'\sum_{ab\sigma}\nonumber \\
&\times G_{ab\sigma}(\tau,\tau')\left(t^2G_{ba\sigma}(\tau',\tau)+\frac{1}{\sqrt{N}}\sum_\alpha V_\alpha^2\bar{\psi}_{\alpha a}(\tau)\psi_{\alpha b}(\tau')\right). \label{eq:Action}
\end{align}
The $\{\bar{c},c\}$ are Grassmann variables and  $a,b=1,\dots,n$ denote replica indices; we have also introduced a large-$N$ field $G_{ab\sigma}(\tau,\tau')=(1/N)\sum_i\bar{c}_{ib}(\tau')c_{ia}(\tau)$. We promote $G$ to a fluctuating field by introducing a Lagrange multiplier field $\Sigma_{ab\sigma}(\tau,\tau')$ and, after integrating out the fermionic fields, obtain $\overline{Z^n}=\int \mc{D}(G,\Sigma)e^{-NS_\mr{eff}[G,\Sigma]}$. 

Assuming a paramagnetic and replica-diagonal symmetric ansatz (i.e.~$G_{aa\sigma}=G$ and $G_{ab\sigma}=0$ for $a\neq b$), we get
\begin{align}
S_\mr{eff}&=\int d\tau d\tau'\left[t^2G(\tau,\tau')G(\tau',\tau)-\Sigma(\tau,\tau')G(\tau',\tau)\right]\nonumber\\
&-\frac{1}{N}\sum_\alpha \mathrm{Tr}\ln(-\mc{G}_\alpha^{-1})-\ln\mc{Z}_\mr{imp},
\end{align}
where 
\begin{align}
\mc{G}^{-1}_\alpha(\tau,\tau')=-(\partial_\tau+\epsilon_\alpha)\delta(\tau-\tau')-V_\alpha^2\sqrt{N} G(\tau,\tau').
 \label{eq.GLIOM} 
\end{align}
We can see that $\mc{Z}_\mr{imp}=\int \mc{D}(\bar{c}_\sigma,c_\sigma)e^{-S_\mr{imp}}$ is the partition function of an effective Anderson impurity model with 
\begin{align}
S_\mr{imp}&=\int d\tau d\tau'\sum_\sigma\bar{c}_\sigma(\tau)\tilde{\mc{G}}^{-1}(\tau,\tau')c_\sigma(\tau')\nonumber \\
&-\int d\tau Un_\uparrow (\tau)n_\downarrow(\tau). \label{eq.ImpurityAction}
\end{align}
Here, $\tilde{\mc{G}}^{-1}(\tau,\tau')=-\partial_\tau\delta(\tau-\tau')-\Sigma(\tau,\tau')$. 

We use a saddle-point approximation to obtain the bubble fermions Green's function self-consistently. To this end, the self-consistency conditions are obtained by setting $\delta S_\mr{eff}/\delta G(\tau,\tau')=\delta S_\mr{eff}/\delta \Sigma(\tau,\tau')=0$ and assuming time-translation invariance, i.e. $G(\tau,\tau')=G(\tau-\tau')$:
\begin{subequations}\label{eq.HubbardAnderson_Saddle}
\begin{align}
G(\tau)&= \langle \bar{c}_\sigma(0)c_\sigma(\tau)\rangle_\mr{imp}, \\
\Sigma(\tau) &=\Sigma_t(\tau)+\Sigma_V(\tau)=t^2G(\tau)+(1/\sqrt{N})\sum_\alpha V_\alpha^2 \mc{G}_\alpha(\tau). \label{eq:SelfEnergy}
\end{align}
\end{subequations}
The averaging $\langle\dots\rangle_\mathrm{imp}$ is carried out using the effective Anderson impurity action from Eq.~\ref{eq.ImpurityAction} where, due to the time translation invariance at the saddle point,  $\tilde{\mc{G}}^{-1}(\omega)=\omega-t^2G(\omega)-\frac{1}{\sqrt{N}}\sum_\alpha V_\alpha^2\mc{G}_\alpha(\omega)$ for the real-frequency argument $\omega+i0^+$. This closes the self-consistency loop as we obtain $\mc{G}_\alpha(\tau)$ from~Eq.\ref{eq.GLIOM}, provided that the impurity Green's function $\langle \bar{c}_\sigma(0)c_\sigma(\tau)\rangle_\mr{imp}$ can be calculated from the impurity action~\ref{eq.ImpurityAction}. $G(\omega)$ and $\mc{G}_\alpha(\omega)~=~\left[\omega+\epsilon_\alpha-V_\alpha^2 \sqrt{N} G(\omega)\right]^{-1}$ are the retarded Green's functions for the bubble fermions and the LIOMs, respectively.

The action in Eq.~\ref{eq.ImpurityAction} is the usual one for the Anderson-Kondo impurity problem and it is routinely encountered in the implementation of single-site DMFT. The impurity problem can be exactly solved either via a Bethe ansatz~\cite{Andrei1983} or by a numerical renormalization group approach~\cite{Wilson1975,Bulla2008}. In our case, for the self-consistent solution of Eq.~\ref{eq.HubbardAnderson_Saddle}, we use the iterated perturbation theory (IPT) method which is expected to work very well at half-filling in both the weak and strong coupling regimes~\cite{Georges1996}.

Using the IPT approximation, we obtain the impurity Green's function from the Dyson equation:
\begin{align}
G^{-1}(\tau,\tau')=   \widetilde{\mathcal{G}}_\mr{H}^{-1}(\tau,\tau')-\Sigma_U(\tau,\tau'),
\end{align}
where $\widetilde{\mc{G}}_\mr{H}^{-1}(\omega) = \widetilde{\mc{G}}^{-1}(\omega)-U/2$ is the Hartree-corrected Green's function at half-filling. The impurity self-energy is approximated by 
\begin{align}
\Sigma_U(\omega)\approx  U/2+\tilde{\Sigma}^{(2)}(\omega). \label{eq:IntSelfEnergy}
\end{align}
The first term on the right is the Hartree shift and the second term corresponds to the second-order self-energy
\begin{align}
\tilde{\Sigma}^{(2)}(\tau,\tau') =  -U^{2}\widetilde{\mathcal{G}}_\mr{H}^{2}(\tau,\tau')\widetilde{\mathcal{G}}_\mr{H}(\tau',\tau).
\end{align}
Using the IPT Green's function we solve the saddle-point equations~\ref{eq.HubbardAnderson_Saddle} iteratively. 

In particular, we solve them numerically and track the evolution of the bubble and LIOM spectral functions, $\rho(\omega)=-(1/\pi)\mr{Im}G(\omega)$ and $\rho_\alpha(\omega)=-(1/\pi)\mr{Im}\mc{G}_\alpha(\omega)$, respectively, as we increase $M$. The results for $\rho(\omega)$ are shown in Fig.~\ref{fig.HubbardAnderson} for $N=30$ and for a wide range of values $M\sim 3-10^4$, until the spectrum has converged at large $M$. We take $U=2t$, to wit, equal to the bandwidth of the non-interacting GOE random matrix. The LIOM energies $|\epsilon_\alpha|\leq W=2t$ with $t=1$ have been sampled from a uniform random distribution. We emphasize that, unless otherwise mentioned, the following results correspond to a single realization of the $\epsilon_\alpha$'s, without disorder averaging.

In Figs.~\ref{fig.HubbardAnderson}(a) and (b) we show the evolution of $\rho(\omega)$ for a localization length $\xi=10$ at a temperature $T=2$ and in the presence of a bubble-LIOM coupling strength $V=0.5$ with the LIOM density $n_L=1$. For a small number of added LIOMs, the DOS has a semicircular form as expected from the non-interacting part of the Hamiltonian. However, at higher energies near the band edge there is an extended tail, due to the effect of the interaction $U$. As we couple more LIOMs, the DOS eventually becomes rugged, inheriting the sharp spectral peaks characterizing the insulator for a fixed disorder realization $\{\epsilon_{\alpha}\}$ of the LIOM energies. We find that, even for a very large number of added LIOMs [Fig.~\ref{fig.HubbardAnderson}(b)], the DOS retains its qualitative features reminiscent of the original interacting bubble. For comparison, we also show the corresponding non-interacting result for large $M$ when the interaction strength in the bubble is set to zero ($U=0$). We see that the interacting and non-interacting DOSs are very different. Hence, our results for the particular coupling strength $V=0.5$ ostensibly agree with the assumption behind the instability argument, namely that the spectral function of the interacting ergodic region remains qualitatively unchanged even when a large number of LIOMs are hybridized with it.

Conversely, the results for the DOS at a larger coupling strength $V=1.5$ are shown in Figs.~\ref{fig.HubbardAnderson}(c) and (d). Note that the \emph{effective} coupling of the bubble with the closest LIOM in this case is still weak, $V/\xi=0.15\ll t,U$, and the coupling to LIOMs farther away decays exponentially with distance. Hence, this regime is well within the purview of the instability argument~\cite{DeRoeck2016}, which treats the coupling perturbatively through the FGR. We find that the DOS of the bubble undergoes a striking change, albeit when a substantially large number $M$ of LIOMs are coupled, entirely destroying the spectrum of the original Hubbard model. The DOS becomes almost entirely dominated by the discrete poles of the localized sites. In fact, as shown in Fig.~\ref{fig.HubbardAnderson}(d), for large $M$ and over an energy interval $-W<\omega<W$ the DOS is identical to that of the system where $U=0$ in the bubble. This implies that the interaction effects become irrelevant due to the feedback of the insulator onto the bubble. On the contrary, the smooth tails at $|\omega|>2$ are not caused by the LIOMs, but by the interaction effects in the Hubbard bubble: note that the tails at $2<|\omega|<4$ are unaffected by coupling to the insulator, i.e. the tails of the $M=0$ spectral function in Fig.~\ref{fig.HubbardAnderson}(a) are the same as the tails of the large-$M$ spectral function in Fig.~\ref{fig.HubbardAnderson}(d).

The above ``strong'' back-action of the insulator onto the bubble can be understood as a consequence of the cumulative, emergent, self-energy effect of a large number of LIOMs, as captured by $\Sigma_V$ (the last term of Eq.~\ref{eq:SelfEnergy}). The contribution of an individual LIOM to the self-energy is $\Sigma_V\sim (V_\alpha/\xi)^2/(W\sqrt{N})$ for $W\apgt t,U$. For $M\gg \pi \xi^2$, namely when the LIOMs over a radius much larger than $\xi$ are added, the typical $\Sigma_V$ is expected to saturate to $\sim V^2/(W\sqrt{N})$, becoming independent of $\xi$ and varying as $\sim 1/\sqrt{N}$. As shown in Fig.~\ref{fig.SelfEnergies}, the strong back-action appears for sufficiently strong $V$, when a large number of LIOMs are coupled to the bubble and $\Sigma_V\apgt \Sigma_t,~\Sigma_U$; the latter two are the self-energies in the DMFT formalism due to the hopping and the interactions in the bubble, respectively.

Figures~\ref{fig.SelfEnergies}(a)--(c) show how the imaginary part of $\Sigma_V$ dominates those of the other two contributions to the self-energy as we increase $M$ for $V=1.5$. We emphasize that the mechanisms giving rise to the competition between self-energies are fundamentally dynamical, as shown by the strong energy ($\omega$) dependence of $\Sigma_t(\omega)$, $\Sigma_V(\omega)$, and $\Sigma_U(\omega)$. Moreover, one self-energy leads to a strong feedback onto the other self-energies via the DMFT self-consistency. 

To better understand the effect of the coupling strength $V$, we compare the typical self-energies, defined as the geometric mean over the interval $-W<\omega<W$ and over several disorder realizations. Figs.~\ref{fig.SelfEnergies}(d) and (e) correspond to $V=1.5$ and $V=0.5$, respectively. In the former case, $\Sigma_V$ becomes comparable to $\Sigma_t$ and $\Sigma_U$ for $M\approx 300$, consistent with the onset of the strong back-action in the spectral function shown in Fig.~\ref{fig.HubbardAnderson}(d). Conversely, for $V=0.5$ shown in Fig.~\ref{fig.SelfEnergies}(e), $\Sigma_V\ll \Sigma_t,\Sigma_U$ and the effect of the LIOMs on the bubble is very weak [Figs.~\ref{fig.HubbardAnderson}(a)--(b)].

This brings us to our main observation. Even though the coupling $V=1.5$ is ``strong'' because it induces a strong back-action, it is \emph{``weak''} in the sense of the ETH theory~\cite{DeRoeck2016} since the effective coupling of the closest LIOM is $(V/\xi)=0.15\ll t=1$. In this regime, the avalanche theory based on ETH would assess the back-action by considering the change of the spectral function upon adding LIOMs one at a time. Within this sequential approach, if the change of spectral function after coupling the closest LIOM is small, then it is expected to remain so after adding farther LIOMs (whose couplings decay exponentially with the distance).  

Indeed, in the strong back-action regime shown in Fig.~\ref{fig.HubbardAnderson}(c), we find that there is no substantial change after coupling a significant number of LIOMs. However, after coupling an even larger number, as shown in Fig.~\ref{fig.HubbardAnderson}(d), the strong back-action on the bubble spectral function becomes apparent. We stress that this emergent effect \emph{cannot} be captured within the ETH framework~\cite{DeRoeck2016}. Ref.~\onlinecite{DeRoeck2016} alludes to certain cumulative strong back-action effects such as an MBL proximity effect \`a la Ref.~\onlinecite{Nandkishore2015}. However, the effect we described is different from an MBL proximity one since it is caused by quantum fluctuations induced by virtual hops of fermions from the Anderson insulator onto the interacting bubble.

Nonetheless, our study of $\rho(\omega)$ in this model cannot indicate whether the system is localized or not---generically, the thermal spectral function of an interacting system at a finite temperature does not contain direct information about localization~\cite{Nandkishore2014a}. Moreover, capturing localization effects in this model might require the inclusion of non-perturbative effects in $1/N$, whereas we have only kept the leading term up to $\mathcal{O}(1/\sqrt{N})$. Nevertheless, the saddle-point results in the effective Anderson impurity model capture the thermal spectral function of the bubble quite accurately for large system sizes. The drastic change of the thermal spectral function $\rho(\omega)$, which we can compute, implies that the spectral functions of typical eigenstates $\rho_{typ}(\omega)$ must also change substantially as a function of $M$, even in the regime of weak bubble-LIOM coupling. This might have important consequences for the critical bubble size, estimated naively from the ETH arguments presented in Section II, as we discuss in the concluding section below.

\section{Discussion and Conclusions}\label{sec:Conclusions}

We have assessed the stability of the MBL phase in two and higher dimensions and in the presence of long range interactions, in light of recent arguments that have called this stability into question~\cite{DeRoeck2016}.  Specifically, it was argued that an instability to rare regions of weak disorder occurring naturally in an insulator can trigger an avalanche that would eventually thermalize the entire system.

As a first test of the assumptions underlying the arguments from Ref.~\onlinecite{DeRoeck2016}, we have used an exact diagonalization study of small systems. The numerical calculations modeled an ergodic bubble coupled to a localized system in both one- and two-dimensional geometries. We found that the numerical results are in excellent agreement with a refined theory of the avalanche based on ETH that also includes corrections due to small system sizes. Thus, our numerical calculations provide further evidence for the validity of the avalanche scenario. The only failure of thermalization in the two-dimensional geometry, as well as in the one-dimensional geometry with stretched exponential interactions, occurred when the ergodic bubble was below a critical size also predicted by the ETH arguments.

As a second test of the avalanche scenario, we have analyzed an effective model of an ergodic bubble coupled to an Anderson insulator. The goal in this analysis has been to check the assumption, central to the avalanche scenario, that the bubble spectral function does not suffer a significant back-action from coupling to the LIOMs. From the numerical solution of the effective model, we found that even for reasonably weak bubble-LIOM coupling there could be substantial back-action of the insulator onto the bubble, leading to a strong modification of the bubble spectral function.

The back-action of the surrounding insulator on the bubble presents a possible mode of failure of the avalanche for a given bubble size. However, we have seen that the effective coupling that generates the back-action is suppressed by $1/\sqrt{N}$ as we increase the bubble size $N$. Thus, the avalanche can always recover from the back-action effect if the bubble is sufficiently large. We conclude that the main effect of the back-action is to renormalize the critical bubble size needed to sustain the avalanche. Hence, the system may well be delocalized as predicted by the avalanche arguments~\cite{DeRoeck2016}, but seeing that it is so will require much larger system sizes than predicted by naive arguments based on ETH.

While our analysis generally lends support to the avalanche scenario, it also highlights the unrealistically large thermalization time scales required to observe the instability, even without the back-action effect. Consider a strongly disordered system, deep in the putative insulating phase, such that its localization length is almost vanishing (i.e. much smaller than a lattice constant). The analysis in Section~\ref{subsec:LIOM_spectral} implies that the minimal bubble size required to sustain an avalanche is $\sim1/\xi \gg 1$. Bubbles of this size represent extremely rare fluctuations occurring with a frequency that decays at least as $\exp(-2A/\xi^2)$ with decreasing $\xi$. In other words, the distance between such bubbles is at least $l_b \sim \exp(A/\xi^2)$. This quickly becomes much larger than any reasonable system size; even if the system is large enough, the time for the thermalization avalanche to reach LIOMs that are not close to any of the bubbles is  
\be
\tau\sim e^{l_b/\xi}= \exp\left[\xi^{-1}e^{A/\xi^2}\right].
\ee
In other words, the local thermalization time quickly becomes very large (note, however, that $\xi$ is related to the logarithm of the disorder strength: $\xi\sim 1/\log W$). Thus, on practical time scales, systems deep in the MBL state remain localized and do not suffer from this instability. On the other hand, the avalanche instability can have a significant effect in the vicinity of the many-body localization transition, eliminating the sharp phase transition. In future work, it would be interesting to explore how the instability, if it indeed occurs, interrupts the critical scaling and broadens the transition into a crossover.

\section*{Acknowledgements}
We would like to thank Dmitry Abanin, Philipp Dumitrescu, Rahul Nandkishore, Wojciech De Roeck, and Maksym Serbyn for many fruitful conversations. We are also grateful to the anonymous referee for the many useful suggestions and comments on our manuscript. This research was supported, in part, by the ERC synergy grant UQUAM. SB acknowledges support from The Infosys Foundation, India.
\appendix

\bibliography{MBL}
\begin{widetext}
\renewcommand{\thesection}{S\arabic{section}}    
\renewcommand{\thefigure}{S\arabic{figure}}
\renewcommand{\theequation}{S\arabic{equation}} 

\setcounter{figure}{0}
\setcounter{equation}{0}
\setcounter{section}{0}

\section{Supplementary Information}

\subsection{Exact diagonalization of a generic model}\label{sec:ED_supp}

The model we are studying is defined as
\begin{equation}
\mathcal{H} = \mathcal{H}_{b} + \mathcal{H}_{l} + \mathcal{H}_{bl}.
\end{equation}
As mentioned in the main text, $\mathcal{H}_{b}$ is a $2^N \times 2^N$ Hermitian matrix sampled from the Gaussian Orthogonal Ensemble such that the many-body bandwidth $\mathcal{W} = E_{\mathrm{max}}-E_{\mathrm{min}}$ (where $E_{\mathrm{max}}$ and $E_{\mathrm{min}}$ are the largest and smallest eigenvalues, respectively) scales linearly with the system size, i.e. $\mathcal{W} = wN$ and $w \approx2.6$. For the remainder of the paper, we place the ergodic quantum dot at the origin and we fix its size to $N=8$ bath spins defined by the Pauli operators $\sigma^{\{x,y,z\}}$. The insulating region consists of $M$ LIOMs defined by the Pauli operators $\tau^{\{x,y,z\}}$ and the Hamiltonian is given by $\mathcal{H}_l = \sum_{\alpha=1}^{M} \epsilon_{\alpha}\tau_{\alpha}^{z}$, where the local fields $\epsilon_{\alpha}$ are sampled from the uniform distribution on $[0.5,1.5]$. Lastly, $\mathcal{H}_{bl} = \sum_{\alpha} V_{\alpha} \sigma_{1}^{x}\tau_{\alpha}^{x}$ describes the bubble-insulator coupling and the $V_{\alpha}$'s depend on the geometry under consideration, as described below.

First, in a $d=2$ geometry with exponentially decaying couplings, we arrange the LIOMs in concentric layers around the ergodic bubble: the $n^{\mathrm{th}}$ layer has a radius $r_{n} = (n-1)a$ (in units of the inter-layer distance $a=1$) and it contains $n$ LIOMs such that the insulator-bubble coupling strength is $V_{\alpha}= V_{1} e^{-r_{n}/\xi}$ for $\alpha =\frac{n(n-1)}{2}+1,...,\frac{n(n+1)}{2}$. We set $V_1=1$ for the first LIOM and $\xi$ is the localization length. The total number of LIOMs, $M$, is related to the total number of layers $n$ via $M = \frac{n(n+1)}{2}$. Second, in a $d=1$ geometry with exponentially decaying couplings we also take $V_{\alpha}= V_{1} e^{-r_{\alpha}/\xi}$, but $r_{\alpha} = (\alpha-1)$ for $\alpha = 1,...,M$. Third, in a $d=1$ geometry with stretched exponentials we take $V_{\alpha}= V_{1} e^{-\sqrt{r_{\alpha}/\xi}}$, where $r_{\alpha} = (\alpha-1)$ and $\alpha = 1,...,M$. 

In all three scenarios, we couple up to $M=6$ LIOMs to the $N=8$ bubble spins and obtain the many-body eigenstates $\ket{\Psi_n}$ and eigenvalues $E_{n}$ of the full Hamiltonian: $\mathcal{H}\ket{\Psi_n}=E_{n}\ket{\Psi_n}$.

\subsubsection{Spectral functions}\label{subsec:Spectral_functions}

In the main text we have defined the spectral function of a local operator $\mathcal{O}$ in an eigenstate $\ket{\Psi_n}$ via 
\begin{equation}
\rho_{n}(\omega) = 2\pi\sum_{m\neq n} |\bra{\Psi_{n}}\mathcal{O}\ket{\Psi_{m}}|^2\delta(\omega-\omega_{mn}),
\end{equation}
where $\omega_{mn} = E_{m}-E_{n}$. Note that $\rho_n(\omega)$ obeys a sum rule whereby $\int_{-\infty}^{+\infty}\rho_{n}(\omega)d\omega  = 2\pi\left(1-\bra{\Psi_n}\mathcal{O}\ket{\Psi_n}^2\right) \approx 2\pi$.

For a local operator $\mathcal{O}= \sigma_{1}^{x}$ acting in the ergodic region, one expects $\rho_{n}(\omega)$ to be a smooth function for a thermal eigenstate and a set of narrow peaks for an MBL eigenstate~\cite{Johri2015}. Moreover, in the MBL case, the narrow peaks occur at different frequencies for different eigenstates even within a given disorder realization. Thus, another diagnostic of localization can be obtained from the ``breakdown of typicality'' of the eigenstate spectral function. To this end, we define the \emph{typical} spectral function via the geometric mean:
\begin{equation}\label{eq:typical_spectral}
\rho_{\mathrm{typ}}(\omega) = \exp\left(\frac{1}{N_A}\sum_{n} \overline{\log \rho_{n}(\omega)}\right),
\end{equation}
where, as before, the overline corresponds to disorder averaging and we also average over $N_A=2^{N+M}/(N+M)$ eigenstates in the middle of the band. For the above equation to be well-defined, we have to consider the limiting cases in which the eigenstate spectral function in a given disorder realization is either zero or a delta function at a frequency $\widetilde{\omega}$. First, if we have $\rho_n(\widetilde{\omega})=0$ then we implicitly take $\rho_{\mathrm{typ}}(\widetilde{\omega})$ to be defined as the limit of the right hand side, namely $\rho_{\mathrm{typ}}(\widetilde{\omega}) = 0$. Second, since we are interested in a finite-size system that has a discrete spectrum, we always work with a finite energy binning and we take the bin size to be $2\delta_{E}$: thus, a delta function peak in a thermodynamic system becomes a narrow peak of width $2\delta_E$ and height $(2\delta_{E})^{-1}$ in a finite-size system. Note that a similar prescription is to introduce an even wider ``energy smearing'' by replacing delta functions with finite-width Lorentzians whose tails give non-zero contributions everywhere. While the two approaches are equally valid, we choose the former since it is computationally faster: see section~\ref{subsec:Numerical_spectral} for details.

Thus, if the system is thermal then $\rho_{\mathrm{typ}}(\omega)$ should be non-zero and a smooth function. However, in contrast to the eigenstate spectral function, $\rho_\mr{typ}(\omega)$ does not have an exact sum rule for $\mathcal{I} = \int_{-\infty}^{\infty}\rho_{\mathrm{typ}}(\omega)d\omega$, but rather an upper bound for $\mathcal{I}$, as detailed in the section below. On the other hand, for an MBL system $\rho_{\mathrm{typ}}(\omega)$ vanishes since $\rho_n(\omega)$ consists of discrete peaks and the peaks occur at different frequencies for different eigenstates even for a single disorder realization.

\subsubsection{Bounds on the typical spectral function}\label{subsec:Bound_typical_spectral}

Since $\rho_n(\omega) \geq 0$ for all frequencies and $n$'s, we can apply the inequality of arithmetic and geometric means to find that
\begin{equation}
0 \leq \rho_{\mathrm{typ}}(\omega) \leq \frac{1}{N_A}\sum_{n} \overline{\rho_{n}(\omega)} = \rho_{\mathrm{th}}(\omega),
\end{equation}
where $\rho_{\mathrm{th}}(\omega)$ is the thermal spectral function at infinite temperature:
\begin{equation}\label{eq:thermal_spectral}
\rho_{\mathrm{th}}(\omega) = \frac{1}{N_A}\sum_{n} \overline{\rho_{n}(\omega)}.
\end{equation}
Defining $\mathcal{I} = \int_{-\infty}^{+\infty}\rho_{\mathrm{typ}}(\omega)d\omega$ we see that
\begin{equation}
0 \leq \mathcal{I} \leq \int_{-\infty}^{+\infty} \rho_{\mathrm{th}}(\omega)d\omega = \frac{1}{N_A}\sum_{n} \overline{\int_{-\infty}^{+\infty}\rho_{n}(\omega)} \approx 2\pi.
\end{equation}
Thus, the upper bound for the typical spectral function's sum rule is $2\pi$. Since the geometric and arithmetic means are equal solely when all numbers being averaged are equal, then $\mathcal{I}$ saturates the $2\pi$ bound if and only if $\rho_{n}^{(i)}(\omega) = \rho_{m}^{(j)}(\omega)$ for any $n,m,i,j$, where $\rho_{n}^{(i)}(\omega)$ is the spectral function in an eigenstate $\ket{\Psi_{n}}$ in the $i^{\mathrm{th}}$ disorder realization. Conversely, $\mathcal{I}$ saturates the $2\pi$ upper bound when the typical and thermal spectral functions coincide, $\rho_\mathrm{typ}(\omega) = \rho_{\mathrm{th}}(\omega)$.

Thus, we expect that in the MBL phase $\mathcal{I} \approx 0$ due to the breakdown of typicality and in the thermal phase $\mathcal{I} \approx 2\pi$.

\subsubsection{Numerical implementation}\label{subsec:Numerical_spectral}

The definition in Eq.~\ref{eq:typical_spectral} is reasonable for a thermodynamic system. However, for a finite system and for a finite number of disorder realizations, $\rho_{\mathrm{typ}}(\omega)$ will be dominated by the rare instances in which $\rho_{n}^{(i)}(\omega) = 0$.  To be more precise, suppose we are interested in the value of $\rho_{\mathrm{typ}}(\omega)$ at a given frequency $\omega = \omega_0$ and we are taking the geometric average over $\mathcal{N}$ eigenstate spectral functions ($\mathcal{N}$ is the product of $N_A$ and the number of disorder realizations). Under the definition in Eq.~\ref{eq:typical_spectral}, if a single number out of these $\mathcal{N}$ values is zero and the remaining $\mathcal{N}-1 \gg 1$ are non-zero then $\rho_{\mathrm{typ}}(\omega_0) = 0$ which runs counter to our intuition behind ``typicality''.

As mentioned before, a solution would be to replace delta functions by finite-width Lorentzians whose tails contribute non-zero values everywhere. Since we found this to be computationally slow, we chose the following alternative: we shift all values by $10^{-20}$, namely $\rho_{n}'(\omega) = \rho_{n}(\omega) + 10^{-20}$. If $p\mathcal{N}$ of these values are zero, $(1-p)\mathcal{N}$ are non-zero, and the geometric average of the non-zero ones is $\rho_0$, then shifting everything by $10^{-20}$ gives, to leading order, $\rho_{\mathrm{typ}}'(\omega_0) \approx 10^{-20p}\rho_0^{1-p}$. If $p=0.01$, then $\rho_{\mathrm{typ}}'(\omega_0) \approx 0.6 \rho_{0}$, i.e. if $1\%$ of the values are zero, then our numerically obtained typical spectral function is of the same order of magnitude as $\rho_0$; if $p=0.1$, then $\rho_{\mathrm{typ}}'(\omega_0) \approx 0.01 \rho_{0}^{0.9}$, i.e. if $10\%$ of the values are zero, then our numerically obtained typical spectral function is two orders of magnitude smaller than $\rho_0$. Thus, roughly speaking, ``typical'' means that $\sim 99\%$ of the eigenstate spectral functions share a given feature of interest.

 \begin{center}
\begin{figure*}
\includegraphics[width=1.0\textwidth]{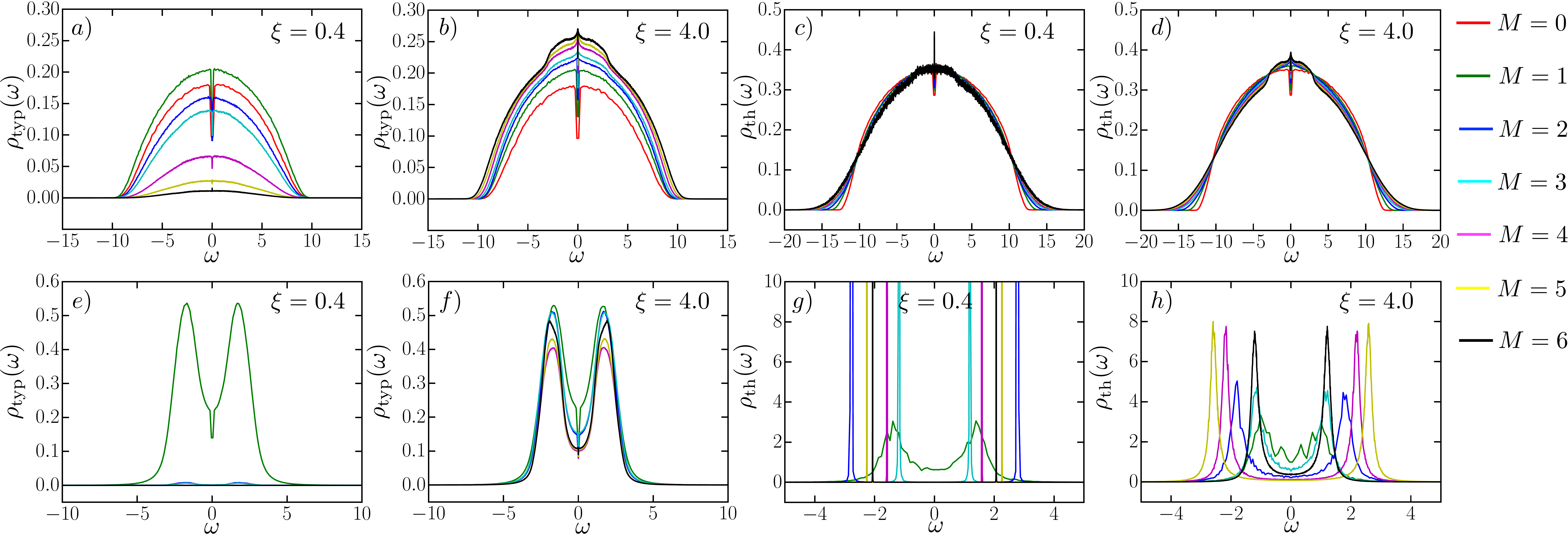}
\caption{(a)--(d) correspond to the spectral functions of a bath spin, whereas (e)--(h) correspond to the spectral functions of a LIOM in a 2D geometry with exponentially decaying couplings. The curves in (a)--(f) have been averaged over $N_A$ eigenstates in the middle of the spectrum and over many disorder realizations. The curves in (g)--(h) correspond to a given disorder realization.}
\label{fig:Spectral_functions}
\end{figure*}
\end{center}

\subsubsection{Bath spectral functions}

For illustrative purposes, in Fig.~\ref{fig:Spectral_functions}(a)--(d) we plot examples of both $\rho_{\mathrm{typ}}(\omega)$ and $\rho_{\mathrm{th}}(\omega)$ for local bath operator $\mathcal{O} = \sigma_{1}^{x}$ in a $d=2$ geometry with exponentially decaying coupling strengths for both small and large localization lengths $\xi$---we note that qualitatively similar behaviors occur for the $d=1$ models with exponentials and stretched exponentials.

In Fig.~\ref{fig:Spectral_functions}(a) we can see the collapse of the typical spectral function for a small localization length, $\xi=0.4$. In Fig.~\ref{fig:Spectral_functions}(c) we plot the \emph{thermal} spectral function for the same parameters and we observe that $\rho_{\mathrm{th}}(\omega)$ is insensitive to the breakdown of typicality. This is due to the fact that $\rho_{\mathrm{th}}(\omega)$ ``washes out'' the differences between the collections of peaks characterizing different eigenstates, leading to a smooth function that remains more or less unchanged with the addition of LIOMs. As emphasized in the main text, this behavior is consistent with the fact that the thermal spectral function \emph{cannot detect localization}. 

We also plot $\rho_{\mathrm{typ}}(\omega)$ and $\rho_{\mathrm{th}}(\omega)$ for a large localization length, $\xi=4.0$ (Fig.~\ref{fig:Spectral_functions}(b) and (d), respectively). First, in both panels we observe the emergence of a plateau at small frequencies for the largest system sizes. This can be understood as an emergent Thouless energy scale~\cite{Serbyn2016}---even though the original ergodic bubble was a zero-dimensional quantum dot, adding spatial structure via the LIOMs gives rise to signatures of locality in the bath. Second, we see that the area under the \emph{typical} spectral function increases monotonically with the addition of LIOMs. 

As argued in Sec.~\ref{subsec:Bound_typical_spectral}, for a thermalizing system, this process will continue until $\rho_{\mathrm{typ}}(\omega) = \rho_{\mathrm{th}}(\omega)$ and $\mathcal{I} \approx 2\pi$. In Fig.~\ref{fig:Integral_Spectral_functions} we further analyze this phenomenon by studying the evolution of $\mathcal{I}$ as a function of the number $M$ of coupled LIOMs for different localization lengths $\xi$. For a large localization length the integral $\mathcal{I}$ increases monotonically with the successive addition of LIOMs, approaching the upper bound allowed by the spectral sum rule. However, in the regime of small localization lengths, the addition of the first few LIOMs strengthens the bubble by increasing $\mathcal{I}$, but coupling more insulating sites eventually collapses the typical spectral function, i.e. $\mathcal{I}\rightarrow 0$. As mentioned in the main text, this breakdown of typicality is a consequence of the fact that the Fermi Golden Rule (FGR) decay rate for the farthest LIOMs is approximately zero.

 \begin{center}
\begin{figure*}
\includegraphics[width=1.0\textwidth]{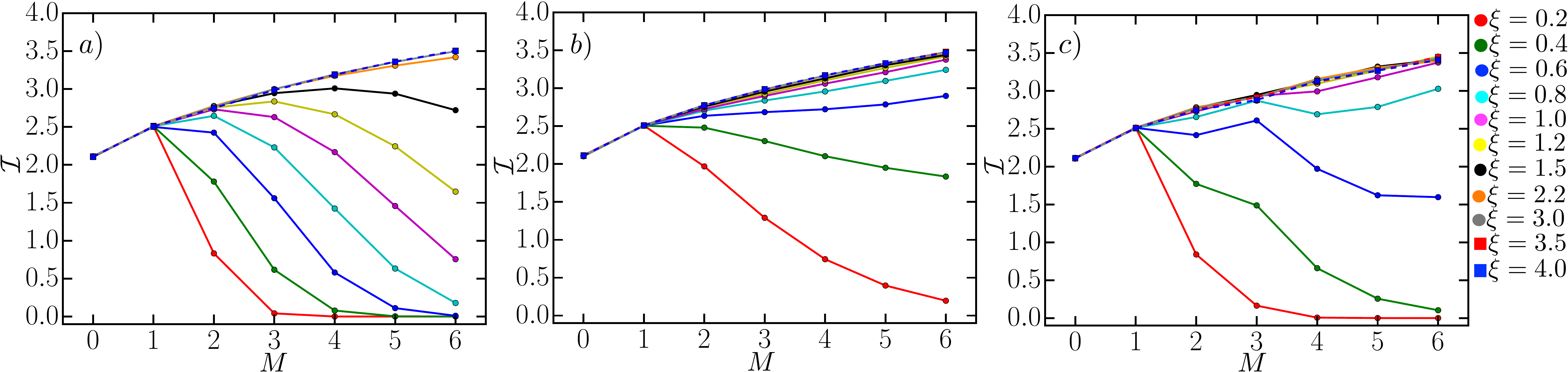}
\caption{The integral over all frequencies for the typical (geometrically averaged over eigenstates and disorder realizations) spectral function of a local operator acting on a bath spin: $\mathcal{I} = \int_{-\infty}^{\infty}\rho_{\mathrm{typ}}(\omega)d\omega$. We plot $\mathcal{I}$ as a function of the LIOMs added and each curve (color) corresponds to a different localization length. (a) Corresponds to a $d=1$ geometry with exponentially decaying couplings. (b) Corresponds to a $d=1$ geometry with stretched exponentials. (c) Corresponds to a $d=2$ geometry with exponentially decaying couplings.}
\label{fig:Integral_Spectral_functions}
\end{figure*}
\end{center}

\subsubsection{LIOM spectral functions}

In the main text we have discussed the width $\gamma$ of the spectral function peaks for a local operator acting on a LIOM, $\mathcal{O} = \tau_{M}^{x}$. We now plot in Figs.~\ref{fig:Spectral_functions}(e) and (f) a few examples of these LIOM spectral functions. In particular, in Figs.~\ref{fig:Spectral_functions}(e) and (f) we plot the typical spectral function $\rho_{\mathrm{typ}}(\omega)$ corresponding the farthest LIOM that was coupled to the ergodic region for $\xi=0.4$ and $\xi=4.0$, respectively. For a small localization length, we also observe the collapse of the \emph{typical} spectral function [Fig.~\ref{fig:Spectral_functions}(e)], whereas for a large localization length we observe two broad peaks [Fig.~\ref{fig:Spectral_functions}(f)] of equal width.

The structure becomes more transparent when we look at the thermal spectral function of the LIOM farthest from the ergodic region in a \emph{fixed disorder realization} for the local fields $\{\epsilon_{\alpha}\}$. As shown in Figs.~\ref{fig:Spectral_functions}(g) and (h), the thermal spectral function corresponds to two sharp peaks located at $\pm 2\epsilon_{M}$, where $1\leq 2|\epsilon_M|\leq 3$. From these we extract the widths $\gamma$ and compute the numerical ratio $\gamma/\delta_E$, as described in the main text, which allows us to check whether the FGR decay rate is non-zero.

\begin{center}
\begin{figure*}
\includegraphics[width=1.0\textwidth]{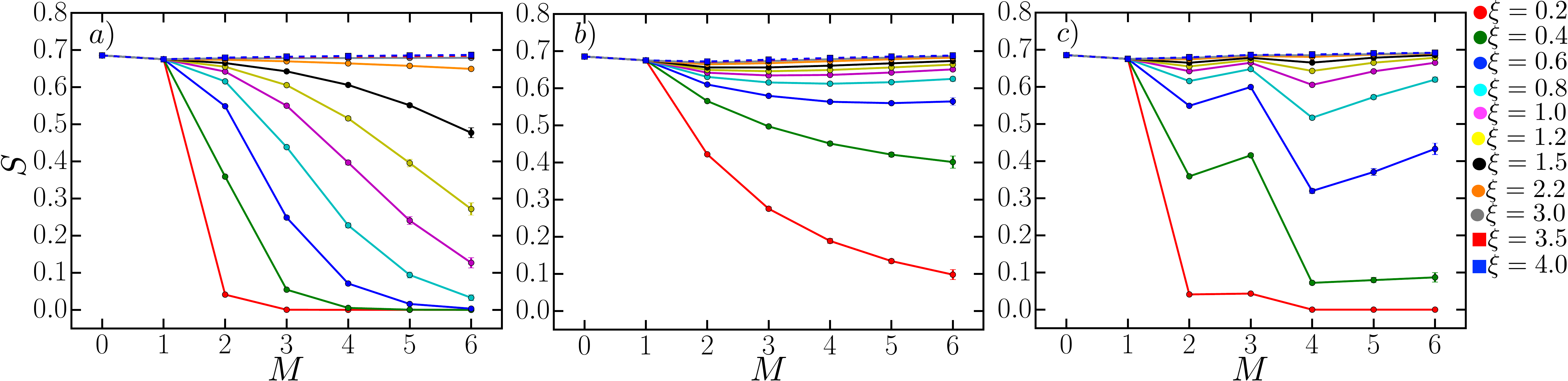}
\caption{The entanglement entropy $S$ of the farthest LIOM coupled to the ergodic bubble as a function of $M$. Each curve (color) corresponds to a different localization length and we have averaged over $N_{A}$ eigenstates in the middle of the spectrum and over many disorder realizations. (a) Corresponds to a $d=1$ geometry with exponentially decaying couplings. (b) Corresponds to a $d=1$ geometry with stretched exponentials. (c) Corresponds to a $d=2$ geometry with exponentially decaying couplings.}
\label{fig:Entropy_SM}
\end{figure*}
\end{center}

\begin{center}
\begin{figure*}
\includegraphics[width=1.0\textwidth]{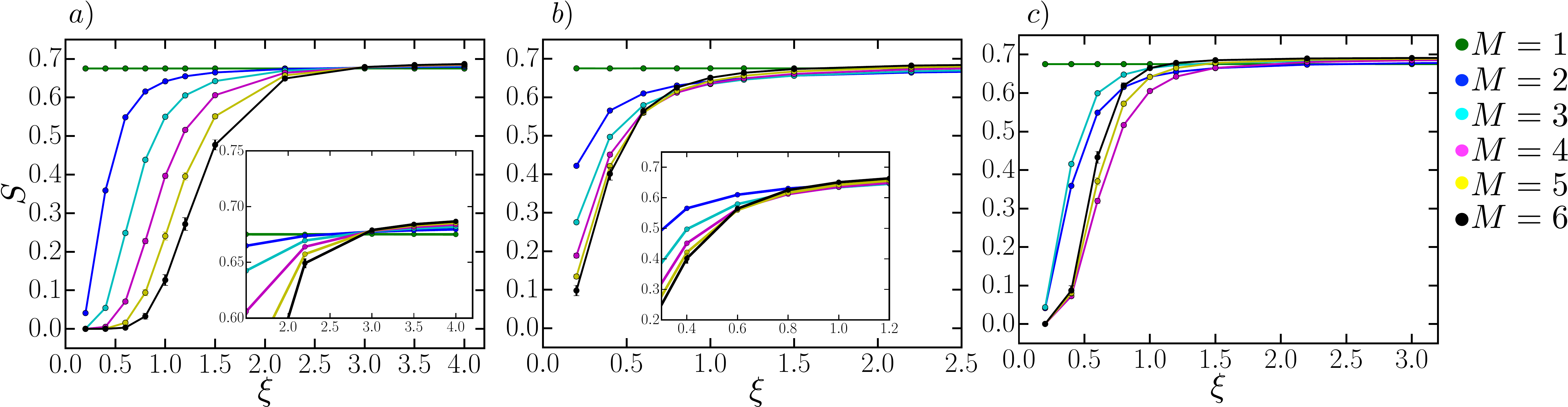}
\caption{The entanglement entropy $S$ of the farthest LIOM coupled to the ergodic region as a function of the localization length $\xi$. Each curve (color) corresponds to a different total number $M$ of LIOMs (i.e. a different system size). (a) Corresponds to a $d=1$ geometry with exponentially decaying couplings. (Inset) A zoom-in around the crossing point $\xi_c \sim 2.9$ between the different curves and this value of $\xi_c$ is in very good agreement with the one obtained in Refs.~\onlinecite{Luitz2017,DeRoeck2016}. (b) Corresponds to a $d=1$ geometry with stretched exponentials. (Inset) A zoom-in around $\xi \sim 0.7$ between the curves, showing that there is no crossing (i.e. it is not a transition). (c) Corresponds to a $d=2$ geometry with exponentially decaying couplings.}
\label{fig:Entropy_Sxi}
\end{figure*}
\end{center}

\subsubsection{Entanglement entropy of the LIOM farthest from the ergodic region}\label{subsec:Entanglement_entropy}

As a final diagnostic, we track the entanglement entropy $S$ of the farthest LIOM coupled to the ergodic bubble. For a many-body eigenstate $\ket{\Psi}$ of the full system we compute the reduced density matrix of the farthest, $M^{\mathrm{th}}$, LIOM: $\rho = \Tr' \ket{\Psi}\bra{\Psi}$, where $\Tr'$ corresponds to tracing out the other $(N+M-1)$ degrees of freedom. Then, the entanglement entropy is defined as $S = -\Tr \left(\rho \log \rho\right)$ and takes a value between 0 (no entanglement) and $\log 2$ (fully entangled). In passing, we have explicitly checked that the entanglement entropy of a bath spin is always the maximal $\log 2$ for all $\xi$'s and $M$'s---this suggests that $\ket{\Psi_{\mathrm{bath}}}$ is a fully thermal state for the $N=8$ spins in the ergodic region regardless of how many LIOMs we couple to it.

As shown in Fig.~\ref{fig:Entropy_SM}, for small enough localization lengths $\xi$ the entanglement entropy eventually collapses, $S\rightarrow 0$ as $M$ increases, indicating that LIOMs far away from the bubble will be disentangled. In the case of a 1D insulator, this is due to the fact that an ergodic bubble, regardless of its initial size, cannot sustain the thermalization avalanche indefinitely for localization lengths $\xi$ below the critical value, $\xi<\xi_c = 2/\log 2$. In the case of a 2D insulator or a 1D insulator with sub-exponentially decaying wave functions, this is due to the fact that the initial bubble size is not large enough to sustain the avalanche: recall that, in the former case, there exists a critical bubble size $N^{*}\sim \xi^{-2}$ below which the avalanche is eventually arrested. Since we fix $N=8$, for small localization lengths, $\xi \lesssim \frac{1}{\sqrt{N}}$, we expect that $S\rightarrow 0$ for large enough $M$. Conversely, for large localization lengths, $\xi \gtrsim \frac{1}{\sqrt{N}}$, we see that $S\rightarrow \log 2$ for large enough $M$ (see Fig.~\ref{fig:Entropy_SM}). 

Secondly, we note that the behavior of $S$ exhibits signatures of the 2D geometry, as shown in Fig.~\ref{fig:Entropy_SM}(a): the entropy slightly increases for LIOMs within a given layer, but it sharply drops as we move on to the next layer. 

Similarly, in Fig.~\ref{fig:Entropy_Sxi} we plot $S(\xi)$ for different $M$'s (system sizes). For the 2D geometry or the 1D geometry with stretched exponentials, we observe a crossover between no entanglement ($S=0$) and maximally entangled ($S=\log2$) as a function of the localization length, as shown in Fig.~\ref{fig:Entropy_Sxi}. For the $d=1$ model with exponentially decaying couplings [Fig.~\ref{fig:Entropy_Sxi}(b)] we find that there is a crossing between the $S(\xi)$ curves corresponding to different system sizes ($M$'s) that occurs at $\xi_c \sim 2.9$. Ref.~\onlinecite{Luitz2017}, which has shown extensive numerical data for this geometry, found that there exists a \emph{transition} at $\xi_{c} = \frac{2}{\log 2} \approx 2.88$ which is in very good agreement with our results.

Lastly, we have also analyzed the average ratio $\overline{r}(\xi) = \min\{\Delta E_{k},\Delta E_{k+1}\}/\max\{\Delta E_{k},\Delta E_{k+1}\}$, where $\Delta E_{k} = E_{k+1} -E_{k}$, and found a behavior very similar to that of $S(\xi)$. Naturally, we find that there is no such crossing and transition for the $d=1$ model with stretched exponentials or the $d=2$ model.

\end{widetext}
\end{document}